\documentclass[twocolumn]{aastex61}

\usepackage{hyperref}
\usepackage{graphicx}
\usepackage{natbib}
\usepackage{amsmath,amsthm,amssymb}
\usepackage{url}
\usepackage{morefloats}

\def\WISE{\textit{WISE}}
\def\WD{\textit{WISE}/DEIMOS}

\protected\def\Lsed{\ifmmode \,\mathcal{L}_{\mathrm{SED}}\else $\mathcal{L}_{\mathrm{SED}}$\fi}

\protected\def\picometer{\ifmmode \,\operatorname{pm}\else $\operatorname{pm}$\fi}
\protected\def\nm{\ifmmode \,\operatorname{nm}\else $\operatorname{nm}$\fi}
\protected\def\micron{\ifmmode \,\operatorname{\mu m}\else $\operatorname{\mu m}$\fi}
\protected\def\mm{\ifmmode \,\operatorname{mm}\else $\operatorname{mm}$\fi}
\protected\def\meter{\ifmmode \,\operatorname{m}\else $\operatorname{m}$\fi}
\protected\def\km{\ifmmode \,\operatorname{km}\else $\operatorname{km}$\fi}
\protected\def\au{\ifmmode \,\operatorname{AU}\else $\operatorname{AU}$\fi}
\protected\def\pc{\ifmmode \,\operatorname{pc}\else $\operatorname{pc}$\fi}
\protected\def\kpc{\ifmmode \,\operatorname{kpc}\else $\operatorname{kpc}$\fi}
\protected\def\Mpc{\ifmmode \,\operatorname{Mpc}\else $\operatorname{Mpc}$\fi}
\protected\def\rsun{\ifmmode \,\operatorname{R_\odot}\else $\operatorname{R_\odot}$\fi}
\protected\def\Rsun{\ifmmode \,\operatorname{R_\odot}\else $\operatorname{R_\odot}$\fi}

\protected\def\second{\ifmmode \,\operatorname{sec}\else $\operatorname{sec}$\fi}
\protected\def\yr{\ifmmode \,\operatorname{yr}\else $\operatorname{yr}$\fi}
\protected\def\Gyr{\ifmmode \,\operatorname{Gyr}\else $\operatorname{Gyr}$\fi}

\protected\def\eV{\ifmmode \,\operatorname{eV}\else $\operatorname{eV}$\fi}
\protected\def\keV{\ifmmode \,\operatorname{keV}\else $\operatorname{keV}$\fi}
\protected\def\MeV{\ifmmode \,\operatorname{MeV}\else $\operatorname{MeV}$\fi}
\protected\def\GeV{\ifmmode \,\operatorname{GeV}\else $\operatorname{GeV}$\fi}
\protected\def\TeV{\ifmmode \,\operatorname{TeV}\else $\operatorname{TeV}$\fi}

\protected\def\Lsun{\ifmmode \,\operatorname{L_\odot}\else $\operatorname{L_\odot}$\fi}
\protected\def\lsun{\ifmmode \,\operatorname{L_\odot}\else $\operatorname{L_\odot}$\fi}
\protected\def\Watt{\ifmmode \,\operatorname{W}\else $\operatorname{W}$\fi}
\protected\def\nW{\ifmmode \,\operatorname{nW}\else $\operatorname{nW}$\fi}

\protected\def\kJy{\ifmmode \,\operatorname{kJy}\else $\operatorname{kJy}$\fi}
\protected\def\Jy{\ifmmode \,\operatorname{Jy}\else $\operatorname{Jy}$\fi}
\protected\def\mJy{\ifmmode \,\operatorname{mJy}\else $\operatorname{mJy}$\fi}
\protected\def\microJy{\ifmmode \,\operatorname{\mu Jy}\else $\operatorname{\mu Jy}$\fi}
\protected\def\nJy{\ifmmode \,\operatorname{nJy}\else $\operatorname{nJy}$\fi}

\protected\def\Mag{\ifmmode \,\operatorname{mag}\else $\operatorname{mag}$\fi}

\protected\def\deg{\ifmmode ^{\circ}\else $^{\circ}$\fi}
\protected\def\arcsec{\ifmmode ^{\prime\prime}\else $^{\prime\prime}$\fi}

\protected\def\arcsecT{\ifmmode \,\operatorname{arcsec}\else $\operatorname{arcsec}$\fi}
\protected\def\arcmin{\ifmmode ^{\prime}\else $^{\prime}$\fi}

\protected\def\arcminT{\ifmmode \,\operatorname{arcmin}\else $\operatorname{arcmin}$\fi}
\protected\def\sr{\ifmmode \,\operatorname{sr}\else $\operatorname{sr}$\fi}

\newcommand{\code}[1]{\texttt{#1}}

\protected\def\d{\ifmmode \operatorname{d}\else
    $\operatorname{d}$\fi}
\protected\def\e{\ifmmode \operatorname{e}\else
    $\operatorname{e}$\fi}

\def\apjl{{ApJ}}

\def\apjs{{ApJS}}

\def\apjref#1;#2;#3;#4 {\par\pp#1, {#2}, #3, #4 \par}

\shorttitle{WISE Galaxy Luminosity Function}
\shortauthors{Lake et al.}

\begin{document}

\title{The 2.4 $\mu$m Galaxy Luminosity Function as Measured Using \WISE. III. Measurement Results}
\author[0000-0002-4528-7637]{S.~E.~Lake}
\affiliation{Physics and Astronomy Department, University of California, Los Angeles, CA 90095-1547}

\author[0000-0001-5058-1593]{E.~L.~Wright}
\affiliation{Physics and Astronomy Department, University of California, Los Angeles, CA 90095-1547}

\author[0000-0002-9508-3667]{R.~J.~Assef}
\affiliation{N\'ucleo de Astronom\'ia de la Facultad de Ingenier\'ia y Ciencias, Universidad Diego Portales, Av. Ej\'ercito Libertador 441, Santiago, Chile}

\author{T.~H.~Jarrett}
\affiliation{Astronomy Department University of Cape Town Private Bag X3 Rondebosch 7701 Republic of South Africa}

\author[0000-0003-0624-3276]{S.~Petty}
\affiliation{NorthWest Research Associates
4118 148th Ave NE
Redmond, WA 98052-5164}

\author{S.~A.~Stanford}
\altaffiliation{Institute of Geophysics and Planetary Physics, Lawrence Livermore National
Laboratory, Livermore CA 94551}
\affiliation{Department of Physics, University of California, Davis, CA 95616}

\author{D.~Stern}
\affiliation{Jet Propulsion Laboratory, California Institute of
Technology, 4800 Oak Grove Dr., Pasadena, CA 91109}

\author[0000-0002-9390-9672]{C.-W.~Tsai}
\altaffiliation{Jet Propulsion Laboratory, California Institute of
Technology, 4800 Oak Grove Dr., Pasadena, CA 91109}
\affiliation{Physics and Astronomy Department, University of California, Los Angeles, CA 90095-1547}


\correspondingauthor{S.~E.~Lake}
\email{lake@physics.ucla.edu}

\begin{abstract}
The \WISE\ satellite surveyed the entire sky multiple times in four infrared wavelengths \citep[3.4, 4.6, 12, and $22\micron$;][]{Wright:2010}. 
The unprecedented combination of coverage area and depth gives us the opportunity to measure the luminosity function of galaxies, one of the fundamental quantities in the study of them, at $2.4 \micron$ to an unparalleled level of formal statistical accuracy in the near infrared. 
The big advantage of measuring luminosity functions at wavelengths in the window $\approx 2$ to $3.5\micron$ is that it correlates more closely to the total stellar mass in galaxies than others. 
In this paper we report on the parameters for the $2.4\micron$ luminosity function of galaxies obtained from applying the spectroluminosity functional based methods defined in \cite{Lake:2017a} to the data sets described in \cite{Lake:2018a} using the mean and covariance of $2.4\micron$ normalized spectral energy distributions (SEDs) from \cite{Lake:2016}. 
In terms of single Schechter function parameters evaluated at the present epoch, the combined result is: {$\phi_\star = 5.8 \pm [0.3_{\mathrm{stat}},\, 0.4_{\mathrm{sys}}] \times 10^{-3} \operatorname{Mpc}^{-3}$}, {$L_\star = 6.4 \pm [0.1_{\mathrm{stat}},\, 0.3_{\mathrm{sys}}] \times 10^{10}\, L_{2.4\micron\,\odot}$} ($M_\star = -21.67 \pm [0.02_{\mathrm{stat}},\, 0.05_{\mathrm{sys}}]\operatorname{AB\ mag}$), and {$\alpha = -1.050 \pm [0.004_{\mathrm{stat}},\, 0.04_{\mathrm{sys}}]$}. 
The high statistical accuracy comes from combining public redshift surveys with the wide coverage from \WISE, and the unevenness in statistical accuracy is a result of our efforts to work around biases of uncertain origin that affect resolved and marginally resolved galaxies. 
With further refinements, the techniques applied in this work promise to advance the study of the spectral energy distribution of the universe.
\end{abstract}

\keywords{galaxies: evolution, galaxies: luminosity function, mass function, galaxies: statistics}


\section{Introduction}
The luminosity function (LF) is one of the most basic statistical properties measured for any class of objects in astronomy. 
The fundamental nature of the LF means that it has been measured for galaxies many times, in many different bandpasses \citep[a small sample: ][]{Bell:2003, Blanton:2003LF, Cool:2012, Dai:2009, Jones:2006, Kelvin:2014, Kochanek:2001, Lin:1999, Loveday:2012, Loveday:2000, Montero:2009, Smith:2009}. 

The release of the AllWISE catalog generated from the data gathered by the \WISE\ satellite, described in \cite{Wright:2010} and \cite{AllWISE}, marks the availability of $3.4 \micron$ (W1) and $4.6 \micron$ (W2) photometric data that is better than 95\% complete over the vast majority of the sky down to $44$ and $88 \microJy$ ($19.79$ and $19.04 \operatorname{AB\ mag}$), respectively. 
This new data set presents the opportunity to utilize the large number of public redshift surveys, in tandem with a small \WISE-selected survey of our own, to measure the near-IR luminosity function of galaxies at $2.4\micron$ to unprecedented accuracy. 
The advantage of measuring the luminosity function in this range of wavelengths is that fluxes suffer from minimal dust extinction in both the target galaxy and the Milky Way, according to dust extinction models like the one from \cite{Cardelli:1989}. 
Further, near infrared light traces the target galaxy's stellar mass in evolved stars more faithfully than optical wavelengths \citep{Loveday:2000}, as long as the contribution of thermally pulsing asymptotic giant branch (TP-AGB) stars can be correctly accounted for in the population synthesis models \citep[for example:][]{Bruzual:2003, Maraston:2005}. 
We measure the luminosity function at $2.4\micron$, in particular, because it is the wavelength directly observed by W1 for galaxies at the median redshift, $z=0.38$, of galaxies with $F_{\mathrm{W1}} > 80\microJy$ in \cite{Lake:2012}.

This new opportunity also presents new challenges.
First, even limiting the redshift surveys to those that are publicly available and that are, primarily, selected by flux at a single wavelength meant that there were a lot of details that needed to be addressed in the characterization and selection process for the six surveys used here. 
Because of this, the primary characterization is done separately in a companion paper \citep[LW18II][]{Lake:2018a}. 
Second, we limited the surveys to be well above the sensitivity limits of the AllWISE data set, a minimum flux of $80\microJy$ in W1 ($19.14$ AB mag), in order to minimize the additional incompleteness from the cross-match and to match the properties of \cite{Lake:2012} as closely as possible. 
This means that the surveys have flux limits at two wavelengths. 
The two flux limits, combined with the wide range of redshifts included in the analysis, $0.01 < z \le 1.0$, meant that the existing luminosity function measurement tools were not adequate to the challenge of analyzing the collected data set. 
For this reason, a new estimator based on analyzing the likelihood of a galaxy's entire spectral energy distribution (SED) is derived in a companion paper \citep[LW17I][]{Lake:2017a}. 
The likelihood estimate used in this work is based on the mean and covariance of SEDs as measured in \cite{Lake:2016}. 

The structure of this paper is as follows: Section~\ref{sec:data} summarizes properties of the data used to measure the $2.4\micron$ luminosity function described in LW18II, Section~\ref{sec:theory} summarizes the estimators used in this paper to perform the measurements, Section~\ref{sec:results} contains the results of the analysis (including comparisons to other measured LFs), Section~\ref{sec:discussion} places the results of this work in context and outlines possible improvements to the methods used, and Section~\ref{sec:conclusion} contains the conclusions drawn from the analysis.

The cosmology used in this paper is based on the WMAP 9 year $\Lambda$CDM cosmology \citep{Hinshaw:2013}\footnote{\url{http://lambda.gsfc.nasa.gov/product/map/dr5/params/lcdm_wmap9.cfm}}, with flatness imposed, yielding: $\Omega_M = 0.2793,\ \Omega_\Lambda = 1 - \Omega_M$, redshift of recombination $z_{\mathrm{recom}} = 1088.16$, and $H_0 = 70 \km \second^{-1} \Mpc^{-1}$ (giving Hubble time $t_H = H_0^{-1} = 13.97\operatorname{Gyr}$, and  Hubble distance $D_H =  c t_H = 4.283 \operatorname{Gpc}$). All magnitudes are in the AB magnitude system, unless otherwise specified. 
In cases where the source data was in Vega magnitudes and a magnitude zero point was provided in the documentation, they were used for conversion to AB (2MASS\footnote{\url{http://www.ipac.caltech.edu/2mass/releases/allsky/faq.html\#jansky}} and AllWISE\footnote{\url{http://wise2.ipac.caltech.edu/docs/release/allsky/expsup/sec4_4h.html\#WISEZMA}}). 
For the surveys without obviously documented zero points (NDWFS\footnote{\url{http://www.noao.edu/noao/noaodeep/}}, SDWFS\footnote{\url{http://irsa.ipac.caltech.edu/data/SPITZER/SDWFS/}}) we converted Vega magnitudes to AB magnitudes using those provided in \cite{AGES}. 
When computing bandpass solar luminosities we utilized the 2000 ASTM Standard Extraterrestrial Spectrum Reference E-490-00\footnote{\url{http://rredc.nrel.gov/solar/spectra/am0/}}. 
For our standard bandpass, W1 at $z=0.38$, we get an absolute magnitude of $M_{2.4\micron\, \odot} = 5.337\operatorname{AB\ mag}$, $L_{2.4\micron\, \odot} = 3.344 \times 10^{-8} \Jy \Mpc^2$.

\section{Data Summary}\label{sec:data}
The full description of the data sets used in this work can be found in LW18II.  
Table~1 of LW18II summarizes the properties of the spectroscopic surveys used here, and we refer interested readers there.

As described in LW17I, the parametric model used here requires models for the relationship between noise and flux. 
Because the flux selection cuts are all relatively bright a single power law of the form
\begin{align}
	\left(\frac{\sigma_F}{F}\right)^2 = \left(\frac{F}{A}\right)^B \label{eqn:photnoisemod}
\end{align}
was adequate. 
The model also requires an estimate of the mean, $\langle \tau\rangle$, and standard deviation, $\sigma_{\tau}$, of the dust obscuration for each selection filter in the field of the redshift survey. 
Table~\ref{tbl:NoiseModels} contains the parameters measured for each survey, with separate values for the optical selection filter and W1. 
The calculation of the incompleteness due to flux variety also uses the parameters in the table to calculate the selection function in redshift-luminosity space, $S(L,z)$.

\floattable
\begin{deluxetable}{l|cccccccc}
	\tabletypesize{\scriptsize}
	\tablewidth{\textwidth}
	\tablecaption{Noise Models Used for Parametric Fits and Completeness}
	\tablehead{ \colhead{Survey} & 
		\colhead{$A_{\mathrm{opt}}$} & \colhead{$B_{\mathrm{opt}}$} & 
		\colhead{$A_{\mathrm{W1}}$} & \colhead{$B_{\mathrm{W1}}$} &
		 \colhead{$\langle \tau_{\mathrm{opt}}\rangle$} & \colhead{$\sigma_{\tau\,\mathrm{opt}}$} & 
		 \colhead{$\langle \tau_{\mathrm{W1}}\rangle$} & \colhead{$\sigma_{\tau\,\mathrm{W1}}$} \\
		  &
		  \colhead{ \nJy } &  & 
		  \colhead{ \nJy } &  &
		  \colhead{$10^2$} & \colhead{$10^2$} & 
		  \colhead{$10^3$} & \colhead{$10^3$} }
	\startdata
		6dFGS & $43410$ & $-0.9068$ & $5.469\times 10^{-6}$ & $-0.2774$ & $2.559$ & $3.142$ & $12.69$ & $15.58$\\
		SDSS & $203.7$ & $-1.174$ & $0.9552$ & $-0.5334$ & $0$ & $0$ & $6.173$ & $6.107$\\
		GAMA & $66.15$ & $-1.419$ & $6.793$ & $-0.6317$ & $0$ & $0$ & $5.914$ & $1.801$\\
		AGES & $37.12$ & $-1.472$ & $1.570$ & $-0.5651$ & $2.002$ & $0.4561$ & $2.015$ & $0.4591$\\
		\WISE/DEIMOS & $89.52$ & $-0.457$ & $675.6$ & $-1.016$ & $23.91$ & $23.89$ & $15.8$ & $15.79$\\
		zCOSMOS & $102.1$ & $-1.379$ & $139.1$ & $-0.8576$ & $3.134$ & $0.2279$ & $3.141$ & $0.2283$\\
	\enddata
	\tablecomments{Noise model parameters fit to each survey separately. 
	The `$\mathrm{opt}$'  parameters pertain to the optical filter used for selection in the survey, 
	and the `$\mathrm{W1}$' parameters are the parameters for the W1 filter. 
	The $A$ and $B$ parameters relate to a power law fit of signal-to-noise ratio to flux (see Equation~\ref{eqn:photnoisemod}), and $\langle \tau\rangle$ and $\sigma_\tau$ are the mean and standard deviation of the optical depth from foreground dust obscuration, averaged over targets in the survey. 
	When the dust obscuration parameters are zero, target selection was performed on fluxes after extinction correction instead of before. 
	All quantities given to four significant figures, regardless of the statistical or systematic uncertainties in the quantities. }
	\label{tbl:NoiseModels}
\end{deluxetable}

\begin{figure*}[htb]
	\begin{center}
	\includegraphics[width=\textwidth]{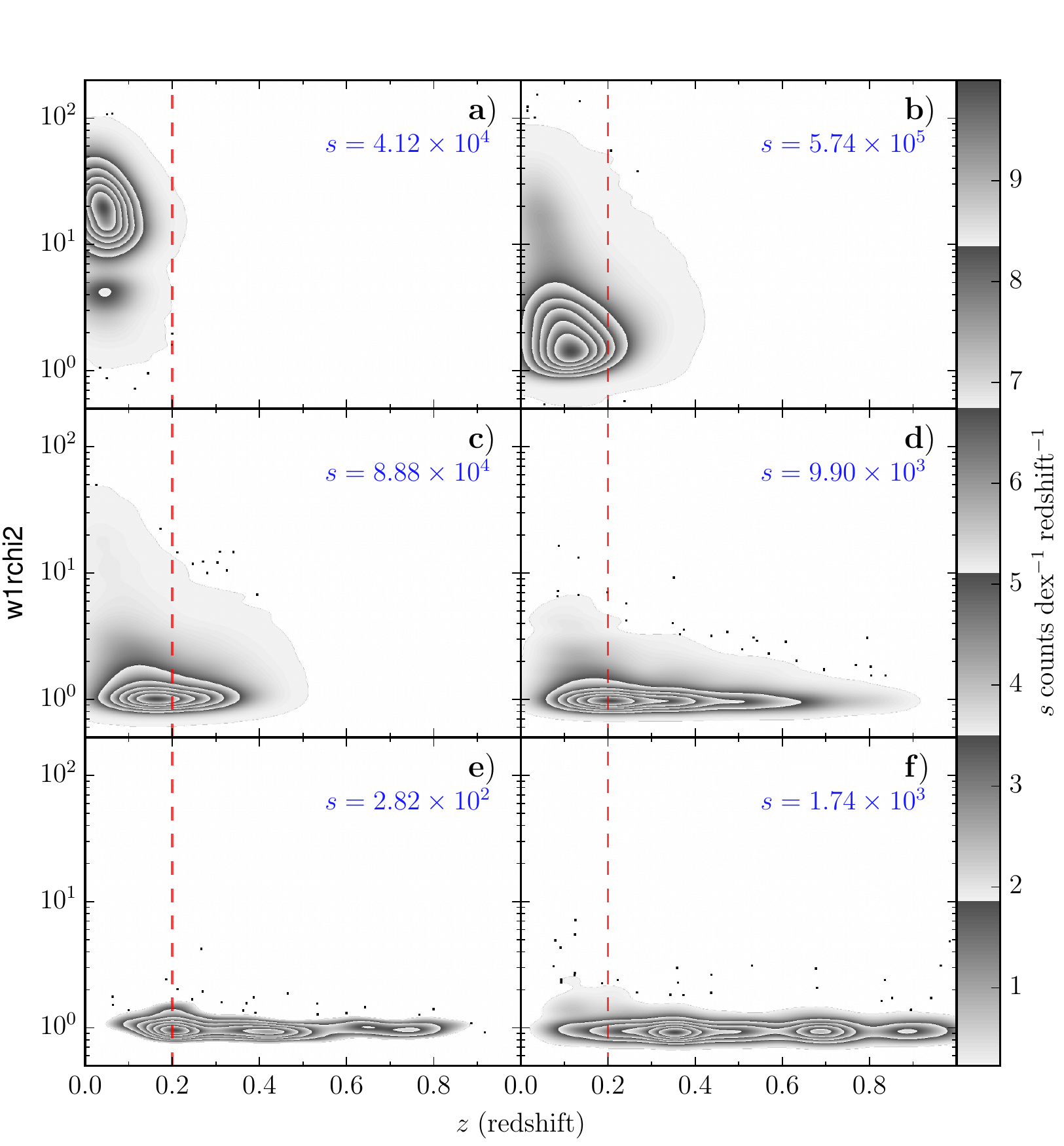}
	\end{center}
	
	\caption{\code{w1rchi2} versus $z$}{Illustration of the trend in \code{w1rchi2} with redshift.
	\code{w1rchi2} is the AllWISE database column containing the reduced $\chi^2$ of fitting the \WISE\ point-spread function to the source and, as long as the source doesn't contain a significant fraction of saturated pixels, is correlated to how resolved a source is.
	Each graph has its own different color bar scale, labeled on the plot as the parameter $s$ (chosen so that the maximum value of each plot is $10\times s$). 
	The vertical line at $z=0.2$ illustrates the dividing line between the low $z$ and high $z$ subsamples. 
	Panel~\textbf{a} is made using the 6dFGS sample, \textbf{b} SDSS, \textbf{c} GAMA, \textbf{d} AGES, \textbf{e} \WD, \textbf{f} zCOSMOS. }
	\label{fig:chivz}
\end{figure*}

There is one additional plot set not in LW18II that is necessary for understanding the choices made about how to analyze the data, found in Figure~\ref{fig:chivz}. 
It shows that \code{w1rchi2}, the reduced $\chi^2$ of fitting a point spread function to the source in the W1 images \citep[for more detailed information, see ][]{AllWISE}\footnote{\url{http://wise2.ipac.caltech.edu/docs/release/allwise/expsup/sec5\_3bii.html\#review\_stationary\_model}},  behaves about as one would expect for galaxies with $z<1$ where, even if galaxies in the past were the same size as today, the increasing angular diameter distance gives them, overall, a smaller radius on the sky. 
The vertical line at $z=0.2$ is placed there after trial and error as a dividing line between samples that are significantly contaminated by resolved and marginally resolved objects ($z \le 0.2$), and those that are sufficiently point-like to render resolution concerns moot ($z > 0.2$). 
In this work we describe the set of all galaxies that come from below $z=0.2$ as the low $z$ sample, and those from above the line as the high $z$ sample. 

As a check on the impact the new analysis techniques from LW17I had on the results, we have also analyzed high and low redshift subsamples that had completeness above $98\%$ of the maximum value for the survey, as defined by Equation~21 of LW17I, and as shown by the light blue contours on the luminosity-redshift plots in LW18II. 
These samples, called the `trim' simples, have significantly lower numbers of galaxies with the accompanying increase in statistical uncertainty, but also reduced systematic uncertainty from the constancy of the selection function. 

The number of sources each survey makes to the combined samples, as well as their overall sizes, can be found in Table~\ref{tbl:combinations}.

\begin{deluxetable*}{l|rrrrrr|r}
	\tabletypesize{\scriptsize}
	\tablecaption{Combined Samples Sizes}
	\tablehead{\colhead{Sample} & \colhead{$N_{\mathrm{6dFGS}}$} & \colhead{$N_{\mathrm{SDSS}}$} &
		\colhead{$N_{\mathrm{GAMA}}$} & \colhead{$N_{\mathrm{AGES}}$} & 
		\colhead{$N_{\mathrm{WD}}$\tablenotemark{a}} & \colhead{$N_{\mathrm{zC}}$\tablenotemark{b}} & \colhead{$N_{\mathrm{tot}}$} }
	\startdata
		Low $z$ & $27,071$ & $450,731$ & $28,619$ & $2,000$ & $36$ & $133$ & $508,590$ \\
		High $z$ & $0$ & $26,556$ & $15,872$ & $3,741$ & $171$ & $1,091$ & $47,431$ \\
		Low $z$ Trim & $15,891$ & $106,003$ & $9,513$ & $1,212$ & $25$ & $116$ & $132,760$ \\
		High $z$ Trim & $0$ & $21$ & $125$ & $452$ & $17$ & $647$ & $1,262$\\
	\enddata
	\tablecomments{Number of sources contributed by each survey to the combined samples named in the first column. 
	The last column contains the total number of sources in each combined sample.}
	\tablenotetext{a}{\WD}
	\tablenotetext{b}{zCOSMOS}
	\label{tbl:combinations}
\end{deluxetable*}

\section{Analysis Methods and Models Summary}\label{sec:theory}
The analyses in this work are centered around maximum likelihood estimates of the quantities related to the LF. 
As is customary in most works on the LF, we analyze the data using both binned and parametric estimators. 
The full description and derivation of the estimators used here is in LW17I. 
Two standard binned estimators were adapted for this work: $1/V_{\mathrm{max}}$ from \cite{Schmidt:1968} as modified by \cite{Avni:1980}; and $N_{\mathrm{obs}}/N_{\mathrm{mdl}}$ from \cite{Miyaji:2001}. 
The adaptation was to handle situations where the selection function, $S(L,z)$, varies significantly over the bin in luminosity-redshift space. Note that the selection function, $S$, is here not regarded as a particular function but as an assessment of the probability that a galaxy with the properties specified in the arguments would be selected for inclusion in the data set.

For the $1/V_{\mathrm{max}}$ estimator, the varying completeness corrected version of the estimator, with log-spaced bins in luminosity, is given by:
\begin{align}
	\Phi(L_i,z_i) = \frac{1}{S(L_i,z_i) L_i \Delta \ln L_i} \sum_{j} \frac{1}{\Delta V_j},
\end{align}
where the redshift-luminosity bin is labeled by the index $i$, $\Phi(L_i,z_i)$ is the constant estimate of the luminosity function for that bin, the sum is over sources that fall into the bin, and $\Delta V_j$ is the volume available to the source to still be in both the bin and the selection criteria of the survey. 

The varying completeness corrected version of $N_{\mathrm{obs}}/N_{\mathrm{mdl}}$ is, formally, much closer to the form of the original estimator from \cite{Miyaji:2001}:
\begin{align}
	\Phi(L_i,z_i) & = \Phi^{\mathrm{mdl}}(L_i,z_i) \cdot \frac{N_i}{\langle N^{\mathrm{mdl}}_i\rangle} \nonumber \\
	& = \frac{\Phi^{\mathrm{mdl}}(L_i,z_i)}{\int_{\mathrm{bin}} \Phi^{\mathrm{mdl}}(L,z)\, S(L,z)\d L \d V} \cdot N_i,
\end{align}
where $\Phi^{\mathrm{mdl}}(L,z)$ is an approximate (`model') luminosity function that brings the estimator closer to evaluating the luminosity function at the center of the bin, $(z_i, L_i)$, as long as $\Phi^{\mathrm{mdl}}(L,z)$ is closer to the true luminosity function than the implicitly assumed $\Phi^{\mathrm{mdl}}(L,z) =$ constant of an uncorrected estimator. 
For these purposes, this work assumes $\Phi^{\mathrm{mdl}}(L,z)$ is a Schechter function that has a faint end slope $\alpha = -1$ and $M_\star = -22\Mag$ (near the peak of the luminosity histograms in LW18II).

The parametric estimator used in this work is based on the spectro-luminosity functional, $\Psi$. 
The final estimator of the likelihood of the data is a little complicated, requiring a few nested equations to express. 
The outermost equation is given by:
\onecolumngrid 
\begin{align}
	\ln(\mathcal{L}) & = \sum_{\mathrm{galaxies}} \ln (S(F_{\mathrm{sel}}, F_0, \vec{x}) \Lsed(F_{\mathrm{sel}}, F_0|L_0) \Phi(L_0, z)) \nonumber \\
		& \hphantom{=} - \int S(F_{\mathrm{sel}}, F_0, \vec{x}) \Lsed(F_{\mathrm{sel}}, F_0|L_0) \Phi(L_0, z) \d F_{\mathrm{sel}} \d F_0 \d L_0 \d V_c, \label{eqn:Likelihood}
\end{align}
\twocolumngrid
\noindent where the selection function, $S$, calculates the probability that a source with given observed fluxes used for primary and secondary target selection, $F_{\mathrm{sel}}$ and $F_0$ (respectively), and at spatial position $\vec{x}$ is selected for inclusion in the survey. 
Here $F_{\mathrm{sel}}$ differs by survey (see the band column of Table~1 in LWII), and $F_0$ is always the W1 flux because the source needs to be well detected there to use $F_{\mathrm{W1}}$ in the calculation of $L_{2.4\micron}$.
This flux selection function is assumed to take only two values: $0$ for excluded regions, and and $s$ for sources in selected regions, where $s$ ($0 < s < 1$) is the overall completeness of the survey. 
Basically, its primary purpose is to set the limits of the integration. 

\Lsed\ is the likelihood for a galaxy to have a particular SED given it has $2.4\micron$ luminosity $L_0$, projected down from the full function space to the fluxes $F_{\mathrm{sel}}$ and $F_0$. 
In this work we approximate the full \Lsed\ as a Gaussian and incorporate an uncertainty model for the fluxes and dust extinction to get:
\onecolumngrid 
\begin{align}
	\vphantom{-}\Lsed(F_{\mathrm{sel}}, F_0|L_0) & = \frac{\e^{\tau_1 + \tau_2}}{\sqrt{(2\pi)^2 \operatorname{det}(\sigma)}} \cdot \left(\frac{4\pi D_L(z)^2}{(1+z)L_0}\right)^2 \cdot \exp \left(-\frac{1}{2} \sum_{i,j=2}^2 [\ell_i - \mu_i][\sigma^{-1}]_{ij}\cdot[\ell_j - \mu_j]\right),\ \mathrm{and} \\
	\ell_i &= \frac{F_i 4\pi D_L(z)^2 \e^{\tau_i}}{(1+z)L_0}. \nonumber
\end{align}
\twocolumngrid
\noindent $\tau_i$ is the optical depth due to dust in the Milky Way present for flux in filter $i$, when the selection fluxes weren't extinction corrected, $0$ when they were. 
$D_L(z)$ is the luminosity distance to the source. The $F_i$ are one of $F_{\mathrm{sel}}$ or $F_0$ and are the measured fluxes. 
$\mu_i$ is the mean value $\ell_i$ takes, as predicted from the mean SED. 
$\sigma$ is a covariance matrix that takes the following form (no summation):
\begin{align}
	\sigma_{ij} = \Sigma_{ij} + \left(\delta_{ij} A_i F_i^{B_i} + \sigma_{\tau\, i} \sigma_{\tau\, j}\right) \mu_i \mu_j,
\end{align}
where $A$ and $B$ are model parameters fit using ordinary least squares in log-space of $\sigma_F^2 F^{-2}$ to $F$, $\sigma_{\tau\, i}$ is the standard deviation of the optical depth present for fluxes in channel $i$ over the survey targets, and $\Sigma_{ij}$ is the covariance of SEDs projected down to apply to the space spanned by $F_{\mathrm{sel}}$ and $F_0$, similar to the calculations described in \cite{Lake:2016}. 
The optical depths are calculated from the dust extinction model of \cite{Cardelli:1989} using the $\operatorname{E}(B-V)$ dust maps from \cite{Schlegel:1998}.

$\Phi(L_0,z)$ is a Schechter luminosity function, originally defined in \cite{Schechter:1976},
\begin{align}
	\Phi(L, z) & = \frac{\phi_\star}{L_\star} \left(\frac{L}{L_\star}\right)^\alpha \e^{-L/L_\star}.
\end{align}
The parameterization for evolution in $\phi_\star$ and $L_\star$ are:
\begin{align}
	\phi_\star & = \phi_0 \e^{-R_\phi t_L(z)},\ \mathrm{and} \nonumber \\
	L_\star & = L_0 \e^{-R_L t_L(z)} \left(1 - \frac{t_L(z)}{t_0}\right)^{n_0}, \label{eqn:evolv}
\end{align}
where $t_L(z)$ is the lookback time of a source at redshift $z$, $R_\phi$ is the specific evolution rate in $\phi$ and is assumed to be constant, $R_L$ is the luminosity evolution rate, $t_0$ is the time of first light (set here to $t_0 = t_L(z_{\mathrm{recom}})$ since our data cannot meaningfully constrain it),  and $n_0$ is the power law index for the early time increase in $L_\star$.

As explained in LW17I, the actual fitting was done using derived parameters that proved to be more statistically orthogonal than the ones given above. 
In terms of the parameters from Equation~\ref{eqn:evolv}, they are:
\begin{align}
	R_n & = R_\phi - \operatorname{min}(1+\alpha, 0) \left[R_L + \frac{n_0}{t_0}\right],\ \mathrm{and} \label{eqn:specnumrate}\\
	\kappa_\star & = \frac{\phi_0 L_0^{3/2}}{\Omega_{\mathrm{sky}} 4\pi^{1/2}} \Gamma\left(\alpha + \frac{5}{2}\right), \label{eqn:kappastar}
\end{align}
which are named the specific rate of change in galaxy number density at $z=0$, and the normalization to the source flux counts distribution. 
If the luminosity function were a static Schechter function in a static Euclidean universe of infinite radius then $\frac{\d N}{\d F \d \Omega} = \kappa_\star F^{-5/2}$.

\subsection{Error Analysis Details}
The statistical uncertainty in the binned estimators is assumed to be fundamentally Poisson combined with propagation of errors that assumes all factors other than the number of galaxies in a bin are constants, as described in LW17I. 
For the parametric estimator in Equation~\ref{eqn:Likelihood}, its complicated form makes performing a second order expansion about the maximum likelihood parameters, as is done in frequentist statistics, to find the uncertainty in those parameters tedious and error prone. 
Further, when there is any region of the parameter space where the likelihood becomes unusually flat, as happens frequently when selection functions are involved, the Taylor expansion must be carried out to higher order to get an estimate in the parameters' uncertainties. 
These conditions make the work done here an ideal case for the application of a Bayesian analysis using Markov Chain Monte Carlo (MCMC) to estimate the uncertainties in the parameters. 
The software tool used to perform this error analysis is the Python package known as \code{emcee} version 2.1.0\footnote{\url{http://dan.iel.fm/emcee/current/}}, described in \cite{emcee}, an implementation of the stretch-move algorithm proposed in \cite{Goodman:2010}.

When doing any MCMC analysis the algorithm needs to run for a number of steps before it starts to provide an accurate and uncorrelated sample of the posterior. 
This process is called `burn in.' 
If the initial point is far from the mode of the posterior, then this first step is dominated by a pseudo-random walk toward that mode, effectively making it an inefficient optimization algorithm. 
This process is even less computationally efficient for emcee because it uses an entire ensemble of independent walkers. 
To short circuit this process, we started the walkers in a ball around an extremum found using the optimize package of SciPy. 
When fitting models as complex as the ones used here, there is the added downside that there are frequently multiple local extrema. 
This factor causes a tradeoff in the usage of emcee and how spread out the initial positions of the walkers are: if the spread is large then the odds of finding a better minimum than the current guess goes up, but walkers will also get stuck in local minima that are uninteresting outliers; if the spread is small, then all of the walkers are characterizing the minimum of interest, but it takes much longer to find any possible lower minima.

In principle, it would be possible to design an algorithm that was mostly emcee, but that periodically trimmed outliers and started new walkers at large distances to search for possible new minima. 
In practice, it is easier to break the process up into different steps. 
In minimum finding mode, the initial spread of the walkers is large, centered on the minimum found by a comparatively efficient algorithm, and it restarts any time a new minimum is found that is lower than the initial one. 
The software then switches to minimum characterizing mode where the initial spread is small, and after a number of burn in steps that are discarded the final sample is produced. 
If a new minimum is found during minimum characterizing mode, then the minimum characterizing process begins again from the beginning, but the switch is not made back to minimum finding mode under the assumption that the improvement from doing so is marginal.

With any Bayesian analysis the prior must be described. 
Table~\ref{tbl:prior1} contains the explicit descriptions of the ranges of the parameters, and the priors assumed on the parameters, for the LF. 
In most cases the priors are flat in the given parameter, and the majority of the remainder are flat in the logarithm of the parameters. 
The exceptions to this are the faint end slopes, $\alpha$, and the initial luminosity index, $n_0$. 
Because $\alpha=-2$ gives an unphysical infinite background radiation, we chose to impose a mildly informative prior with a beta distribution shape that excluded the end points of the allowed interval. 
With $n_0$ the buildup of $L_\star$ should neither be discontinuously fast, so $n_0 > 0$, nor should it be much slower than the integral of a linear accretion rate, so $n_0 < 10$. 
Further, the data used here does not constrain the buildup of $L_\star$ directly, so an informative prior that constrains the value of $n_0$ is used. 
Explicitly, it is expected that their initial luminosity is proportional to the gas accretion rate, and should therefore be reasonably near linear, so we impose $n_0 = 1 \pm 0.2$ with a log-normal distribution.

\begin{deluxetable}{cccccc}
	\tabletypesize{\scriptsize}
	\tablewidth{0.5\textwidth}
	\tablecaption{LF Parameter Priors}
	\tablehead{ \colhead{Param} & \colhead{Min} & \colhead{Max} & \colhead{Units} & \colhead{Prior} & \colhead{Notes} }
	\startdata
		$\kappa_\star$ & $10^{-3}$ & $10^2$ & $\operatorname{Jy}^{3/2}\sr^{-1}$ & $\kappa_\star^{-1}$ & \tablenotemark{a} \\
		$R_{n0}$ & $-50 $ & $50 $ & $t_H^{-1}$ & flat & \tablenotemark{b}\\
		$L_0$ & $10^9 $ & $10^{13} $ & $L_{2.4\micron\, \odot}$ & $L_0^{-1}$ & \tablenotemark{c}  \\
		$R_L$ & $-50$ & $50 $ & $t_H^{-1}$ & flat & \\
		$\alpha$ & $-2$ & $0$ & --- & $-\alpha(2+\alpha)$ & \\
		$n_0$ & $10^{-15}$ & $10$ & --- & $n_0^{-1} \operatorname{e}^{-25 \ln(n_0)^2 / 2}$ & \tablenotemark{d}\\
	\enddata
	\tablecomments{Parameter (Param) ranges (from Min to Max) and priors used in fitting the luminosity function. 
	Priors are given in unnormalized form.}
	\tablenotetext{a}{ Normalization to the static Euclidean number counts. 
	Its relationship to standard Schechter function parameters can be found in Equation~\ref{eqn:kappastar}.}
	\tablenotetext{b}{ Specific rate of change of the galaxy number density at $z=0$. 
	Its relationship to standard Schechter function parameters can be found in Equation~\ref{eqn:specnumrate}.}
	\tablenotetext{c}{ $L_{2.4\micron\, \odot} = 33.44 \nJy \Mpc^2 \leftrightarrow M_{2.4\micron\, \odot} = 5.337 \Mag$ AB absolute.  }
	\tablenotetext{d}{ This prior has mean 1 and standard deviation 1/5. }
	\label{tbl:prior1}
\end{deluxetable}

\section{Luminosity Functions}\label{sec:results}
Running \code{emcee} produces a sequence of model parameters that are distributed as though they were samples taken from the posterior distribution. 
The sequence of parameters is, therefore, the result of the analysis from which all other results are derived. 
The chains are published with this paper and at \url{www.figshare.com} under the digital object identifier (DOI) \href{https://dx.doi.org/10.6084/m9.figshare.4109625}{\code{10.6084/m9.figshare.4109625}} in \href{https://www.gnu.org/software/gzip/}{gzipped} \href{http://irsa.ipac.caltech.edu/applications/DDGEN/Doc/ipac_tbl.html}{IPAC table} format. 
An few example lines from one of the chains can be found in Table~\ref{tbl:chainExample}.

\begin{deluxetable*}{rrcccccc}
	\tabletypesize{\scriptsize}
	\tablewidth{0.97\textwidth}
	\setlength{\tabcolsep}{5.0pt}
	\tablecaption{Example Lines from MCMC Chains}
	\tablehead{  \colhead{\code{StepNum}} & \colhead{\code{WalkerNum}} &
			\colhead{\code{ln\_KappaStar}} & \colhead{\code{R\_n}}  & \colhead{\code{ln\_Lstar}} &  \colhead{\code{R\_L}} & \colhead{\code{alpha}} & \colhead{\code{ln\_n0}} \\
		\colhead{---} & \colhead{---}  &
			\colhead{$\ln \left(\Jy^{3/2} \sr^{-1}\right)$}  & \colhead{$\Gyr^{-1}$} & \colhead{$\ln\left(\Jy \Mpc^2\right)$} & \colhead{$\Gyr^{-1}$} & \colhead{---} & \colhead{---} }
	\startdata
		$0$   &     $0$  &     $2.0414$  &  $ -0.056289$ &  $8.7190$ & $-0.13463$ &   $-1.5071$ &   $-0.30514$ \\   
		$0$    &    $1$   &    $1.8404$   &  $-0.013686$ &  $8.2423$ & $-0.19569$  &  $-1.2721$ &   $-0.03764$  \\
		$0$     &   $2$    &   $2.1309$   &  $\hphantom{-}0.224666$ &  $7.6358$ &  $-0.29812$ &  $-0.9931$  & $-0.01672$   \\
		$0$      &  $3$   &    $2.5518$   &  $\hphantom{-}0.095079$ &  $8.6992$  & $-0.22876$  & $-1.4027$ &   $\hphantom{-}0.41558$   \\
		$0$    &    $4$   &    $1.8430$   &  $-0.007250$ &  $8.2185$ &  $-0.21643$ &   $-1.3473$ &   $\hphantom{-}0.03394$   \\
	\enddata
	\tablecomments{Example lines from one of the chains produced by \code{emcee} in the tables under DOI \href{https://dx.doi.org/10.6084/m9.figshare.4109625}{\code{10.6084/m9.figshare.4109625}}. 
	Floating point values truncated here for brevity, but not in the downloadable tables. 
	\code{StepNum} is the zero indexed step number that the ensemble was at in the chain, and \code{WalkerNum} is the number of the walker which was at the position defined by the row for that step. 
	\code{ln\_KappaStar} is the natural logarithm of $\kappa_\star$ in $\Jy^{3/2} \sr^{-1}$ (see Equation~\ref{eqn:kappastar}). 
	\code{R\_n} is the specific rate of change of the galaxy number density evaluated at the present time in $\Gyr^{-1}$ (see Equation~\ref{eqn:specnumrate}). 
	\code{ln\_Lstar} is the natural logarithm of $L_\star$ in $\Jy \Mpc^2$ evaluated at $z=0$ $\left( M_\star = -\frac{2.5}{\ln 10} \ln \left[\frac{L_\star}{3631\Jy 4\pi 10^{-10} \Mpc^2 }\right]\approx -1.086\ln \left[\frac{L_\star}{\Jy \Mpc^2 }\right] -13.352 \mathrm{\ for\ AB\ absolute\ mag}\right)$. 
	\code{R\_L} is the long term decay constant in $L_\star$ in $\Gyr^{-1}$. 
	\code{alpha} is the faint end slope of the luminosity function. \code{ln\_n0} is the natural logarithm of the early time power law index of the evolution of $L_\star$ (see Equation~\ref{eqn:evolv}) }
	\label{tbl:chainExample}
\end{deluxetable*}

\begin{figure*}[htb]
	\begin{center}
	\includegraphics[width=\textwidth]{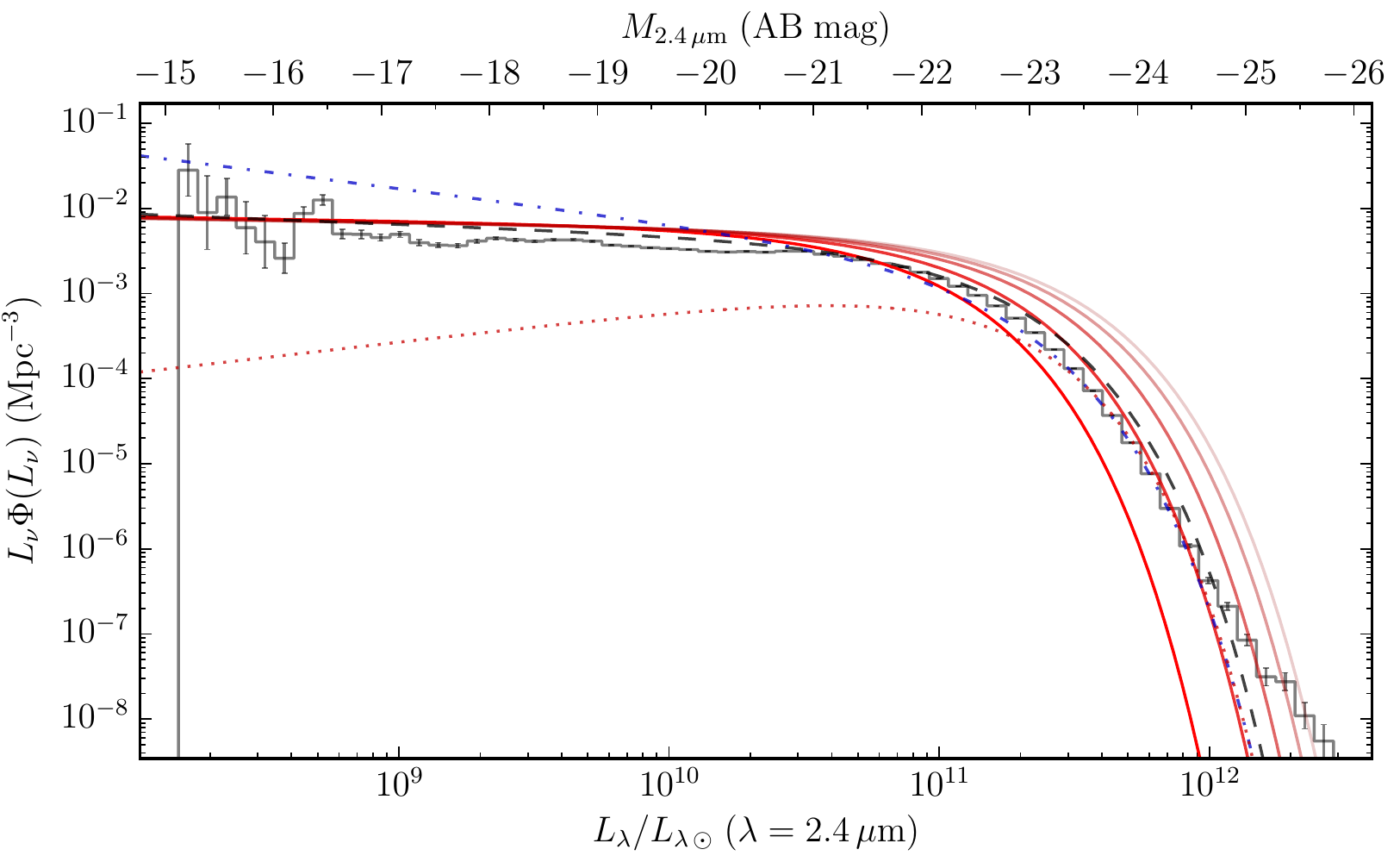}
	\end{center}
	
	\caption{Combined Luminosity Function}{Binned $N_{\mathrm{obs}} / N_{\mathrm{mdl}}$ estimate of the LF compared with fit LFs. Broad agreement among the estimators is apparent, with details emerging that the parametric model does not capture. 
	The grey histogram with error bars is the $N_{\mathrm{obs}} / N_{\mathrm{mdl}}$ estimator applied to the combination of all data. 
	The solid red lines of varying opacity and darkness are the LFs from the mean parameters of the High $z$ Prior chain (see Table~\ref{tbl:res:Parameters}) evaluated at equally spaced redshifts between $z=0.01$ and $1.0$ ($0.01$, $0.258$, $0.505$, $0.743$, $1$), inclusively, with the opacity decreasing as $z$ increases. 
	The remaining lines are based on $3.6\micron$ LF fits from \cite{Dai:2009} evaluated at $z=0.38$ and adapted to $2.4\micron$ using the mean SED from \cite{Lake:2016}, with the black dashed line corresponding to their `all' sample, the red dotted line to their `early' sample, and the blue dash-dotted line to their `late' sample. }
	\label{fig:combinedLF}
\end{figure*}

The content of that file set is as follows: there is an individual chain for each individual spectroscopic survey (with file names matching the names of the survey), one for each merged subsample described in Table~\ref{tbl:combinations} with matching names, and two chains, labeled `High $z$ Prior' and `High $z$ Trim Prior', where the High $z$ samples were analyzed with the mean and standard deviation in $\alpha$ from the matching Low $z$ samples used as additional Gaussian priors in the analysis. 
The reason for only using the faint end slope as a prior is discussed in Section~\ref{sec:internalcomp}. 
Of all of the analysis chains produced, High $z$ Prior is the canonical one for this work based on a subjective evaluation of statistical accuracy and bias. 
A comparison of the $N_{\mathrm{obs}} / N_{\mathrm{mdl}}$ binned estimator using all of the data to the LF with the mean parameters from the High $z$ Prior chain can be found in Figure~\ref{fig:combinedLF}. 
The solid red lines, of varying opacity and brightness, are the result of evaluating the High $z$ Prior LF model at five equally spaced redshifts from $0.01$ to $1$ ($0.01$, $0.258$, $0.505$, $0.743$, $1$). 
The three $3.6\micron$ LFs from \cite{Dai:2009}, color corrected to $2.4\micron$ using the mean SED from \cite{Lake:2016}, are plotted at $z=0.38$ using black dashed (`all'), red dotted (`early'), and blue dash-dotted (`late') lines.

Two features stick out most prominently in Figure~\ref{fig:combinedLF}. 
First, the parametric estimator places the value of $\phi_\star$ a factor of about $1.7$ higher than the binned estimator, though this is within the statistical uncertainty in $\phi_\star$ for Hi $z$ Prior. 
Second, the falloff at the bright end appears to be better described as a power law than exponential. 
The former is likely caused by the fact that the data content of Figure~\ref{fig:combinedLF} is drawn from all the samples, without restriction on redshift, while the plotted LF is of the mean parameters from the High $z$ Prior chain. 
The cause of the latter is uncertain. 
Averaging the evolving fit LF over different redshifts using the observed redshift distribution did not produce the observed power law shape. 
Further, limiting the plot to data in the High $z$ sample does not alter this feature significantly, either, so it is unlikely to be an artifact related to photometry of resolved sources. 
The explanation that, qualitatively, seems most likely to cause the feature is the presence of active galactic nuclei (AGN) in the sample, which are observed to have a power law falloff to the LF on the bright end \citep[see Equation~5, and surrounding discussion, in][]{Richards:2006}. 
While this means that the fit LF doesn't match all of the details of the real LF, this was an expected consequence of using a single Schechter function for the LF and not separate LFs for different galaxy types. 
The global properties of the LF, like predictions of galaxy count and luminosity density, should still correspond with the observable values in the same way that a Gaussian fit to a data set will reproduce the mean and standard deviation, even if the data are not Gaussian distributed.

Detailed analyses of the different posterior chains is done in Subsection~\ref{sec:internalcomp}, including an examination of the necessity of splitting the combined samples by redshift, and evidence for luminosity uncertainty contributing to the softening of the high luminosity falloff. 
Comparisons with the results of other measurements of the luminosity function is done in Subsection~\ref{sec:externalcomp}.

\subsection{Internal Comparisons}\label{sec:internalcomp}
Because the parametric estimator used in this work is new, it is important to analyze multiple data set that all have different selection criteria and compare the results to see if the systematic biases have been correctly managed. 
The mean parameters from each of the posterior chains can be found in Table~\ref{tbl:res:Parameters}, and parameters derived from those mean parameters are found in Table~\ref{tbl:res:modParameters}. 

All of the uncertainties given in the tables are purely statistical uncertainties derived from the Bayesian posterior of the data. 
They do not include sources of error that are, for the purposes of this work, systematic, including: the accuracy of the cosmological parameters ($\approx 3\%$), the peculiar velocity the Milky Way \mbox{($\approx 0.6\%$)}, the accuracy of the completeness assessments of the different surveys \mbox{($\approx 2\%$)}, selection effects not modeled, the accuracy of the AllWISE W1 photometric zero point \citep[$1.5\%$,][]{Jarrett:2011}, and the accuracy of the numerical integration algorithms used \mbox{($\approx 2\%$)}. 
Cosmic variance, the additional variability of the data set induced by the tendency of galaxies to cluster more than random chance, was estimated to be about 5\% for SDSS in \cite{Driver:2010}, and \cite{Driver:2011} found GAMA to be 15\% under-dense with respect to SDSS. Based on this, we assign cosmic variances of: 15\%, 5\%, 15\%, 20\%, 20\%, and 20\% to the 6dFGS, SDSS, GAMA, AGES, \WD, and zCOSMOS surveys, respectively, with 4\% for the combined analyses.
The combination of these effects implies $5.3\%$ systematic uncertainty in the determination of $L_\star$, $5.2\%$ in $\kappa_\star$, $7.3\%$ in $\phi_\star$, an assumed $\e$-fold per Hubble time in the evolution rate parameters, and an assumed $4\%$ in $\alpha$, for the combined samples.

All of the parameters on a given line in Tables~\ref{tbl:res:Parameters} and \ref{tbl:res:modParameters} are correlated, to greater or lesser degrees. 
Including tables or plots of the correlation among the parameters would take up a prohibitive amount of room, so this work only contains a single example of the covariance matrix among the primary parameters constructed from the High $z$ Prior chain in Table~\ref{tbl:res:covar}. 
It should be noted that the full correlation among parameters is not necessarily encapsulated by a covariance matrix, particularly the correlation between $\ln n_0$ and $R_n$. 
In other words, an examination of the pairwise distribution of parameters in each Markov chain does not always show the elliptical structure that would suggest they are well characterized by a Gaussian distribution, and hence by only a mean vector and covariance matrix, particularly when the sample size is small.

The biggest trend in Table~\ref{tbl:res:Parameters} is in $R_n$, with the lower depth surveys consistent with the number density of galaxies currently declining and the higher depth surveys with the opposite. 
While it is possible that this is a real feature of the data, whether from a turnover in the comoving number density of galaxies or cosmic variance from the Milky way existing in an low density region, selection biases must first be ruled out. 
A preliminary analysis of the SDSS subset with a $\mathrm{W1}$ flux maximum of $2\mJy\, (15.65\operatorname{AB\ mag})$ suggested that this effect was being driven by the galaxies with high apparent flux, and not the faint galaxies that dominate number counts for samples with minimum luminosity significantly lower than $L_\star$. 
Because faint galaxies dominate the number density of galaxies, it is unlikely that a trend driven by the presence of bright objects in the sample is a real phenomenon, neither cosmic trend nor cosmic variance.

As the error bars on $R_n$ show, this parameter is poorly constrained by the data, so a small bias in a correlated parameter, even a weekly correlated one, can drive a big change in $R_n$. 
Since bright galaxies are also more likely to be resolved or marginally resolved, we decided to work around the problem by dividing the combined samples at $z=0.2$ where the number of galaxies with high \code{w1rchi2} fell to a level low enough to be negligible (see Figure~\ref{fig:chivz}). 
The advantage of a redshift split instead of a flux cut is that it does not increase the impact of the systematic uncertainties inherent in the the final estimator used. 
The only parameter from the untrimmed low redshift sample analysis that is plausibly not affected by a bias affecting resolved sources is the faint end slope. 
Given the luminosity range covered by the Low $z$ samples makes it the strongest constraint on $\alpha$ available, we use the mean and standard deviation of $\alpha$ from the corresponding Low $z$ samples as Gaussian priors on the High $z$ Prior samples. 

In order to estimate the possible systematic impact the Trim samples were constructed. 
The Trim samples are identical to their corresponding combined samples, but each survey is limited to the region in luminoisty-redshift space where the completeness is at least $98\%$ of its maximum value; the regions enclosed by the faint blue lines in the $L$-$z$ plots of LW18II. 
This substantially reduces the size and depth of the sample, so it is more vulnerable to cosmic and statistical variance. 
The Low $z$ and Low $z$ Trim chains agree within the statistical uncertainties, with the exception of $L_\star$. 
Even though the effects on the High $z$ samples was more dramatic, this is to be expected given the reduced effective depth and loss of low luminosity sources, especially in the AGES and \WD\ samples. 
When the power law part of the LF is not directly sampled, where the LF is linear in a log-log plot, the information about $\alpha$ is encoded in the higher order moments of what was observed, increasing sensitivity to statistical and cosmic variance fluctuations. 

\begin{deluxetable*}{lcccccc} 
	\tabletypesize{\scriptsize}
	\tablewidth{\textwidth}
	\tablecaption{Luminosity Function Bayesian Mean Parameters}
	\tablehead{ \colhead{Survey}  & \colhead{$\kappa_\star $} & \colhead{$R_n$\tablenotemark{a}}  & \colhead{$L_{\star}\tablenotemark{a}$} &  \colhead{$R_L$} & \colhead{$\alpha$} & \colhead{$n_0$} \\
		\colhead{---}  & \colhead{$\Jy^{3/2} \sr^{-1}$} & \colhead{$t_H^{-1}$} & \colhead{$10^{10}\,L_{2.4\micron \, \odot} h^{-2}$} & \colhead{$t_H^{-1}$} & \colhead{---} & \colhead{---} }
	\startdata
		6dFGS & $2.94 \pm 0.07$ & $-4.9 \pm 0.8$ & $3.15 \pm 0.07$ & $-3.4 \pm 0.3$ & $-0.91 \pm 0.02$ & $0.9 \pm 0.2$\\
		SDSS & $4.06 \pm 0.03$ & $-3.55 \pm 0.09$ & $3.34 \pm 0.03$ & $-2.4 \pm 0.2$ & $-0.957 \pm 0.005$ & $1.0 \pm 0.2$\\
		GAMA & $4.0 \pm 0.1$ & $\hphantom{-}0.2 \pm 0.2$ & $2.67 \pm 0.09$ & $-12.7 \pm 0.8$ & $-0.95 \pm 0.02$ & $8.3 \pm 0.6$\\
		AGES & $5.4 \pm 0.3$ & $\hphantom{-}1.8 \pm 0.3$ & $2.2 \pm 0.2$ & $-5.0 \pm 0.7$ & $-0.57 \pm 0.06$ & $1.8 \pm 0.4$\\
		\WISE/DEIMOS & $8 \pm 2$ & $\hphantom{-}1 \pm 2$ & $5 \pm 2$ & $-3 \pm 1$ & $-1.1 \pm 0.3$ & $1.0 \pm 0.2$\\
		zCOSMOS & $4.4 \pm 0.5$ & $\hphantom{-}1.8 \pm 0.8$ & $2.6 \pm 0.5$ & $-4.5 \pm 0.5$ & $-0.9 \pm 0.1$ & $0.9 \pm 0.2$\\
		\hline
		Low $z$ & $3.11 \pm 0.02$ & $-7.47 \pm 0.09$ & $3.41 \pm 0.02$ & $-1.5 \pm 0.2$ & $-1.059 \pm 0.004$ & $0.8 \pm 0.2$\\
		High $z$ & $5.0 \pm 0.2$ & $\hphantom{-}0.6 \pm 0.2$ & $2.66 \pm 0.05$ & $-2.8 \pm 0.1$ & $-0.68 \pm 0.03$ & $0.56 \pm 0.08$\\
		High $z$ Prior & $5.7 \pm 0.2$ & $\hphantom{-}0.1 \pm 0.2$ & $3.12 \pm 0.05$ & $-2.6 \pm 0.1$ & $-1.050 \pm 0.004$ & $0.50 \pm 0.07$\\
		Low $z$ Trim & $3.24 \pm 0.03$ & $-7.7 \pm 0.2$ & $3.62 \pm 0.04$ & $-0.4 \pm 0.2$ & $-0.972 \pm 0.008$ & $0.9 \pm 0.2$\\
		High $z$ Trim & $9 \pm 1$ & $-0.8 \pm 0.7$ & $3.9 \pm 0.4$ & $-4.8 \pm 0.5$ & $-1.93 \pm 0.04$ & $1.0 \pm 0.2$\\
		Hi $z$ Trim Prior & $5.0 \pm 0.6$ & $\hphantom{-}1.9 \pm 0.7$ & $2.9 \pm 0.2$ & $-4.3 \pm 0.4$ & $-0.935 \pm 0.008$ & $1.0 \pm 0.2$\\
	\enddata
	\tablecomments{Mean parameters from the Bayesian posterior functions. 
	The top half of the table is broken down by survey, and the bottom half is one of the combined analyses of all data sets. 
	$\kappa_\star$ is the Euclidean flux counts normalization (see Equation~\ref{eqn:kappastar}). 
	$R_n$ is the specific rate of change of the numeric density of galaxies (see Equation~\ref{eqn:specnumrate}). 
	$R_L$ is the long time decay constant in $L_\star$, and $n_0$ is the initial luminosity index (see Equation~\ref{eqn:evolv}). 
	For $\kappa_\star$, $L_\star$, and $n_0$ the means are geometric means, in keeping with the values published in the posterior chains. }
	\tablenotetext{a}{ Parameter evaluated at $z=0$. }
	\label{tbl:res:Parameters}
\end{deluxetable*}

\floattable
\begin{deluxetable}{lccccccc} 
	\tabletypesize{\scriptsize}
	\tablewidth{\textwidth}
	\tablecaption{Luminosity Function Derived Parameters}
	\tablehead{  \colhead{Survey}  & \colhead{$\phi_\star$\tablenotemark{a}} & \colhead{$R_\phi$} & \colhead{$M_\star\tablenotemark{a} - 5 \log_{10} h $} & \colhead{$z_\star$} & \colhead{$\rho_{L2.4\micron}$\tablenotemark{a}} & \colhead{$R_\rho$\tablenotemark{a}} & \colhead{$z_\rho$} \\
		\colhead{---}  & \colhead{$10^{-2}h^3 \Mpc^{-3}$} & \colhead{$t_H^{-1}$} & \colhead{AB mag} & \colhead{---} & \colhead{$10^8 L_{2.4\micron\,\odot} \Mpc^{-3}$}  & \colhead{$t_H^{-1}$} & \colhead{---} }
	\startdata
		6dFGS & $0.86 \pm 0.03$ & $-5.2 \pm 0.7$ & $-20.91 \pm 0.03$ & $\hphantom{-}1.7 \pm 0.3$ & $1.81 \pm 0.04$ & $-7.6 \pm 0.6$ & $4.1 \pm 0.6$\\
		SDSS & $1.09 \pm 0.01$ & $-3.61 \pm 0.09$ & $-20.973 \pm 0.008$ & $\hphantom{-}1.0 \pm 0.1$ & $2.49 \pm 0.01$ & $-4.93 \pm 0.07$ & $2.8 \pm 0.4$\\
		GAMA & $1.51 \pm 0.07$ & $\hphantom{-}0.0 \pm 0.2$ & $-20.73 \pm 0.04$ & $\hphantom{-}0.43 \pm 0.01$ & $2.76 \pm 0.05$ & $-4.2 \pm 0.2$ & $0.43 \pm 0.02$\\
		AGES & $2.4 \pm 0.2$ & $\hphantom{-}0.4 \pm 0.2$ & $-20.53 \pm 0.09$ & $\hphantom{-}1.3 \pm 0.2$ & $3.4 \pm 0.2$ & $-2.8 \pm 0.4$ & $1.1 \pm 0.1$\\
		\WISE/DEIMOS & $1.2 \pm 0.8$ & $\hphantom{-}1 \pm 2$ & $-21.4 \pm 0.5$ & $\hphantom{-}1.3 \pm 0.6$ & $5 \pm 1$ & $-1 \pm 1$ & $0.5 \pm 0.7$\\
		zCOSMOS & $1.8 \pm 0.5$ & $\hphantom{-}1.5 \pm 0.5$ & $-20.7 \pm 0.2$ & $\hphantom{-}2.3 \pm 0.3$ & $3.1 \pm 0.3$ & $-2.0 \pm 0.4$ & $1.5 \pm 0.3$\\
		\hline
		Low $z$ & $0.814 \pm 0.009$ & $-7.43 \pm 0.09$ & $-20.994 \pm 0.008$ & $\hphantom{-}0.65 \pm 0.09$ & $2.013 \pm 0.009$ & $-8.09 \pm 0.06$ & $4.8 \pm 0.6$\\
		High $z$ & $1.81 \pm 0.08$ & $-0.1 \pm 0.2$ & $-20.73 \pm 0.02$ & $\hphantom{-}2.4 \pm 0.2$ & $3.0 \pm 0.1$ & $-2.4 \pm 0.2$ & $2.5 \pm 0.3$\\
		High $z$ Prior & $1.69 \pm 0.08$ & $\hphantom{-}0.2 \pm 0.2$ & $-20.90 \pm 0.02$ & $\hphantom{-}2.5 \pm 0.3$ & $3.8 \pm 0.1$ & $-1.9 \pm 0.2$ & $2.3 \pm 0.3$\\
		Low $z$ Trim & $0.77 \pm 0.01$ & $-7.7 \pm 0.2$ & $-21.06 \pm 0.01$ & $-0.7 \pm 0.2$ & $1.92 \pm 0.02$ & $-7.1 \pm 0.2$ & $4.2 \pm 0.6$\\
		High $z$ Trim & $1.1 \pm 0.3$ & $\hphantom{-}3 \pm 1$ & $-21.2 \pm 0.1$ & $\hphantom{-}2.4 \pm 0.4$ & $50 \pm 40$ & $-1.1 \pm 0.6$ & $0.8 \pm 0.5$\\
		Hi $z$ Trim Prior & $1.7 \pm 0.3$ & $\hphantom{-}1.7 \pm 0.7$ & $-20.80 \pm 0.08$ & $\hphantom{-}2.1 \pm 0.3$ & $3.3 \pm 0.4$ & $-1.6 \pm 0.5$ & $1.2 \pm 0.3$\\
	\enddata
	\tablecomments{Bayesian mean values related to the luminosity function calculated from the posterior chains separately from the parameters in Table~\ref{tbl:res:Parameters}, with $h=0.7$. 
	The means of $\phi_\star$ and $\rho_{L2.4\micron}$ are geometric means. 
	$R_\phi$ is the specific rate of change of $\phi_\star$ (assumed constant for all redshifts, see Equations~31 of LW17I). 
	$z_\star$ is the redshift at which the model predicts $L_\star$ will peak (Equation~32 of LW17I). 
	$\rho_{2.4\micron}$ is the present day $2.4\micron$ luminosity density of galaxies (see Equation~41 of LW17I), $R_\rho$ is its specific rate of change (Equation~42 of LW17I), and $z_\rho$ is the redshift at which the model predicts $j$ to have peaked (Equation~45 of LW17I). }
	\tablenotetext{a}{ Parameter evaluated at $z=0$. }
	\label{tbl:res:modParameters}
\end{deluxetable}

\begin{deluxetable*}{ll|llllll}
	\tabletypesize{\scriptsize}
	\tablewidth{0.95\textwidth}
	\tablecaption{Hi $z$ Sample with Low $z$ Prior Luminosity Function Bayesian Parameter Posterior Covariance}
	\tablehead{ \colhead{Parameter}  & \colhead{$\sigma$} & \colhead{$\delta \ln \kappa_\star$}  & \colhead{$\delta R_n t_H$} &  \colhead{$\delta \ln L_\star(0)$} & \colhead{$\delta R_L t_H$} & \colhead{$\delta \alpha$} & \colhead{$\delta \ln n_0$} }
	\startdata
		$\delta \ln \kappa_\star$ & $0.03029$ & $\hphantom{-}1.000$ & $\hphantom{-}0.9023$ & $-0.5915$ & $-0.1510$ & $-0.04423$ & $-0.2330$\\
		$\delta R_n t_H$ & $0.2040$ & $\hphantom{-}0.9023$ & $\hphantom{-}1.000$ & $-0.8257$ & $-0.4389$ & $\hphantom{-}0.06990$ & $-0.02677$\\
		$\delta \ln L_\star(0)$ & $0.01524$ & $-0.5915$ & $-0.8257$ & $\hphantom{-}1.000$ & $\hphantom{-}0.7285$ & $-0.1243$ & $-0.2852$\\
		$\delta R_L t_H$ & $0.1210$ & $-0.1510$ & $-0.4389$ & $\hphantom{-}0.7285$ & $\hphantom{-}1.000$ & $-0.009374$ & $-0.8508$\\
		$\delta \alpha$ & $0.004266$ & $-0.04423$ & $\hphantom{-}0.06990$ & $-0.1243$ & $-0.009374$ & $\hphantom{-}1.000$ & $-0.009308$\\
		$\delta \ln n_0$ & $0.1431$ & $-0.2330$ & $-0.02677$ & $-0.2852$ & $-0.8508$ & $-0.009308$ & $\hphantom{-}1.0000$ \\
	\enddata
	\tablecomments{The $\sigma$ column contains the standard deviation of the parameters, and the remaining rows and columns comprise the matrix of correlation coefficients between the parameters, considered pairwise. 
	All values given to four significant figures.}
	\label{tbl:res:covar}
\end{deluxetable*}

\begin{figure*}[p]
	\begin{center}
	\includegraphics[width=0.9\textwidth]{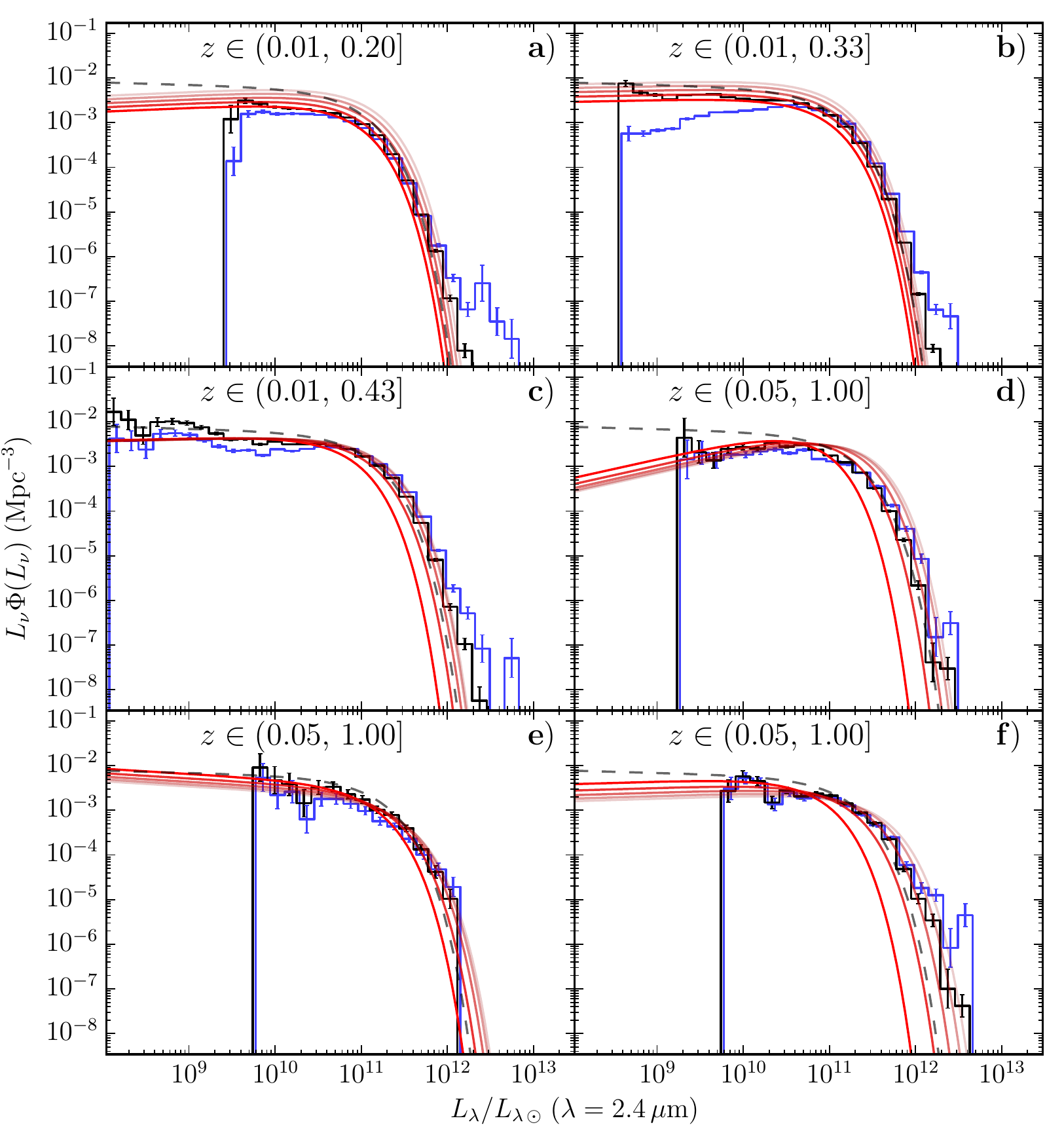}
	\end{center}
	
	\caption{Luminosity Functions}{Above is a grid of plots comparing luminosity function fits (lines) to binned estimators (steps) for the luminosity function over the entire redshift range noted in the panel. 
	The solid red lines are the evolving model LFs fit to the data used to make the panel (not to histograms in the panel), and the dashed line is the mean LF of the High $z$ Prior chain. 
	The combined LFs are evaluated at the middle redshift of the interval in the panel, and the red lines are evaluated at $5$ equally spaced redshifts in their panel's redshift interval (inclusive of endpoints). 
	The black histogram is from the $N^{\mathrm{obs}} / N^{\mathrm{mdl}}$ estimator, and the blue is the $1/V_{\mathrm{max}}$ offset to the right for clarity. 
	Panels \textbf{a} through \textbf{f} are from the surveys: 6dFGS, SDSS, GAMA, AGES, \WD, and zCOSMOS, respectively.}
	\label{fig:global}
\end{figure*}

\begin{figure*}[p]
	\begin{center}
	\includegraphics[width=\textwidth,angle=90]{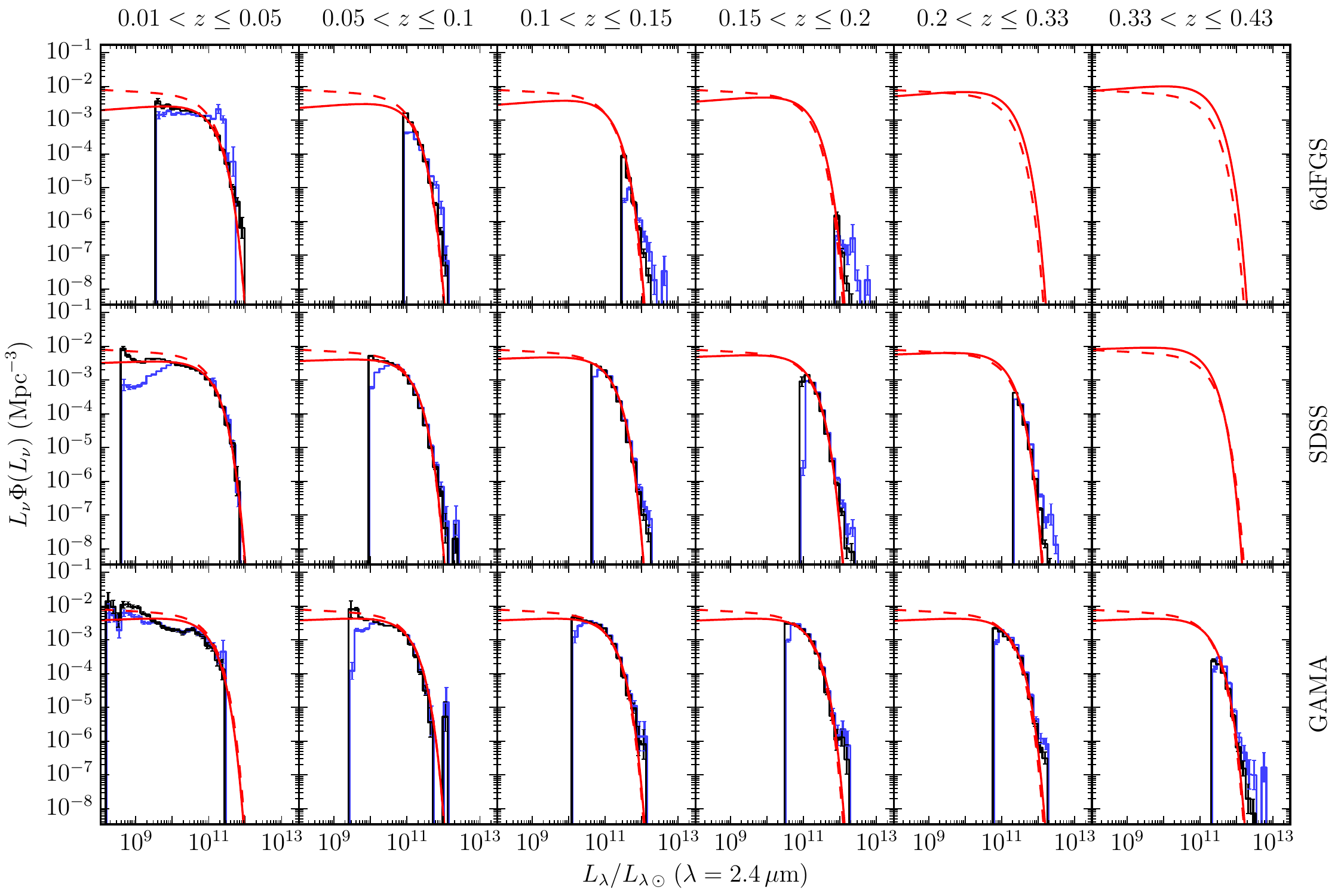}
	\end{center}
	
	\caption{Shallow Evolving Luminosity Functions}{Above is a grid of plots comparing luminosity function fits (lines) to binned estimators (steps) for the shallow surveys. 
	Each column represents a different redshift range noted at its top, and each row corresponds to a different shallow survey noted at its right. 
	The solid lines are the evolving model LFs fit to the entire survey in the row (not to histograms in the row) and the dashed line is the mean LF of the High $z$ Prior chain. 
	All LF lines are evaluated at the middle redshift of the interval in the column.
	The black histogram is from the $N^{\mathrm{obs}} / N^{\mathrm{mdl}}$ estimator, and the blue is the $1/V_{\mathrm{max}}$ offset to the right for clarity.}
	\label{fig:shallow}
\end{figure*}

\begin{figure*}[p]
	\begin{center}
	\includegraphics[width=\textwidth,angle=90]{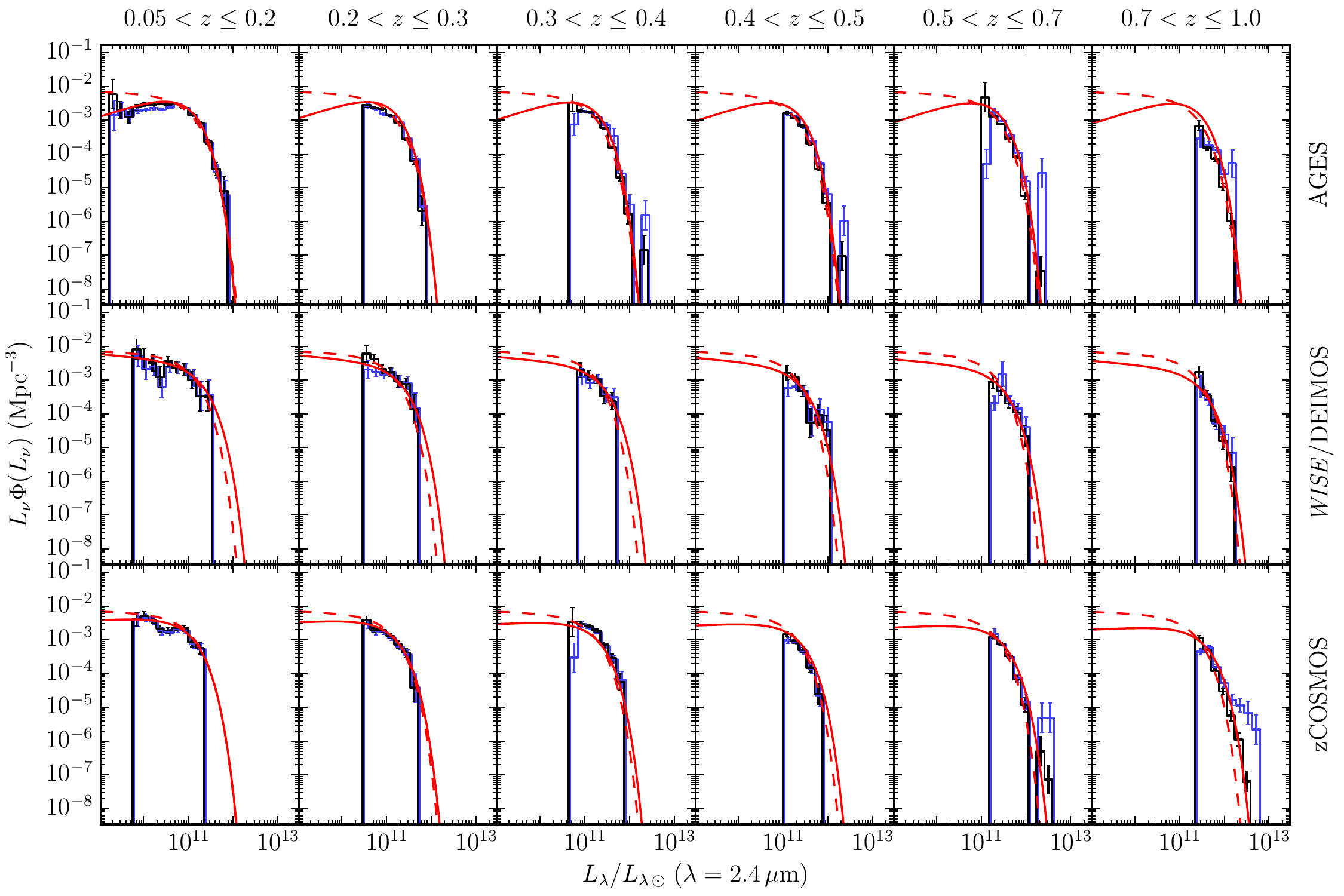}
	\end{center}
	
	\caption{Deep Evolving Luminosity Functions}{Above is a grid of plots comparing luminosity function fits (lines) to binned estimators (steps) for the deep surveys. 
	Each column represents a different redshift range noted at its top, and each row corresponds to a different deep survey noted at its right. 
	The solid lines are the evolving model LFs fit to the entire survey in the row (not to histograms in the row) and the dashed line is the mean LF of the High $z$ Prior chain. 
	All LF lines are evaluated at the middle redshift of the interval in the column.
	The black histogram is from the $N^{\mathrm{obs}} / N^{\mathrm{mdl}}$ estimator, and the blue is the $1/V_{\mathrm{max}}$ offset to the right for clarity.}
	\label{fig:deep}
\end{figure*}

The numerical comparisons discussed above provide a nice overview of the behavior of the different sets, but no work is complete without graphical comparisons of the fit models to binned estimators for the data. 
The comparisons of the unbinned evolving model ML fits to the entire samples with two binned estimators, $1/V_{\mathrm{max}}$ and $N^{\mathrm{obs}} / N^{\mathrm{mdl}}$, can be found in Figures \ref{fig:global}, \ref{fig:shallow}, and \ref{fig:deep}. 
The first notable feature of the grid of plots in Figure~\ref{fig:global} is that the $N^{\mathrm{obs}}/N^{\mathrm{mdl}}$ agrees better with the fit luminosity functions than the $1/V_{\mathrm{max}}$ estimator. 
This is to be expected since $N^{\mathrm{obs}}/N^{\mathrm{mdl}}$ is closer to a maximum likelihood estimator, and the fits were done using unbinned maximum likelihood. 
In particular, the $N^{\mathrm{obs}}/N^{\mathrm{mdl}}$ outperforms $1/V_{\mathrm{max}}$ where the approximation that the selection function is constant with redshift is a bad one. 
The difference is particularly stark for panel \textbf{b}, SDSS, where the shallow optical selection makes SED variability particularly relevant. 
The next obvious feature is the disagreement of the fitted faint end slope with the binned ones in panel \textbf{d}, AGES. 
There is a local maximum in the likelihood with a faint end slope closer to $\alpha = -1$, but it is not a global maximum. 
There are two reasons this fails to be a global maximum: the low faint end completeness of AGES giving the less luminous galaxy bins bigger error bars, and a fluctuation in the data. 
These facts are more apparent in an examination of the first row of Figure~\ref{fig:deep}, where the apparent disagreement vanishes.

The final feature of note in the luminosity function plots is the upturns at the faint ends of panels \textbf{b} (SDSS) and \textbf{c} (GAMA). 
It is likely the same feature that caused \cite{Loveday:2012} and \cite{Kelvin:2014} to use a double luminosity function to fit the data. 
While a double LF would provide better agreement to the data, it is unclear without a deeper examination of the data the extent to which the additional LF is modeling a fundamental feature of the universe (for example, the split between red and blue galaxies) or a cosmic variance fluctuation in the data. 
One example of an even bigger fluctuation can be seen in panel~\textbf{c} of Figure~12 of LW18II. 
There is a significant over-density in the Sloan data near its peak at around $z=0.75$. 
The over-density that causes that bump goes by the name the Sloan Great Wall, discovered in \cite{Gott:2005}, and it is a good example of how large cosmic variance can get.

\begin{figure*}[p]
	\begin{center}
	\includegraphics[width=0.9\textwidth]{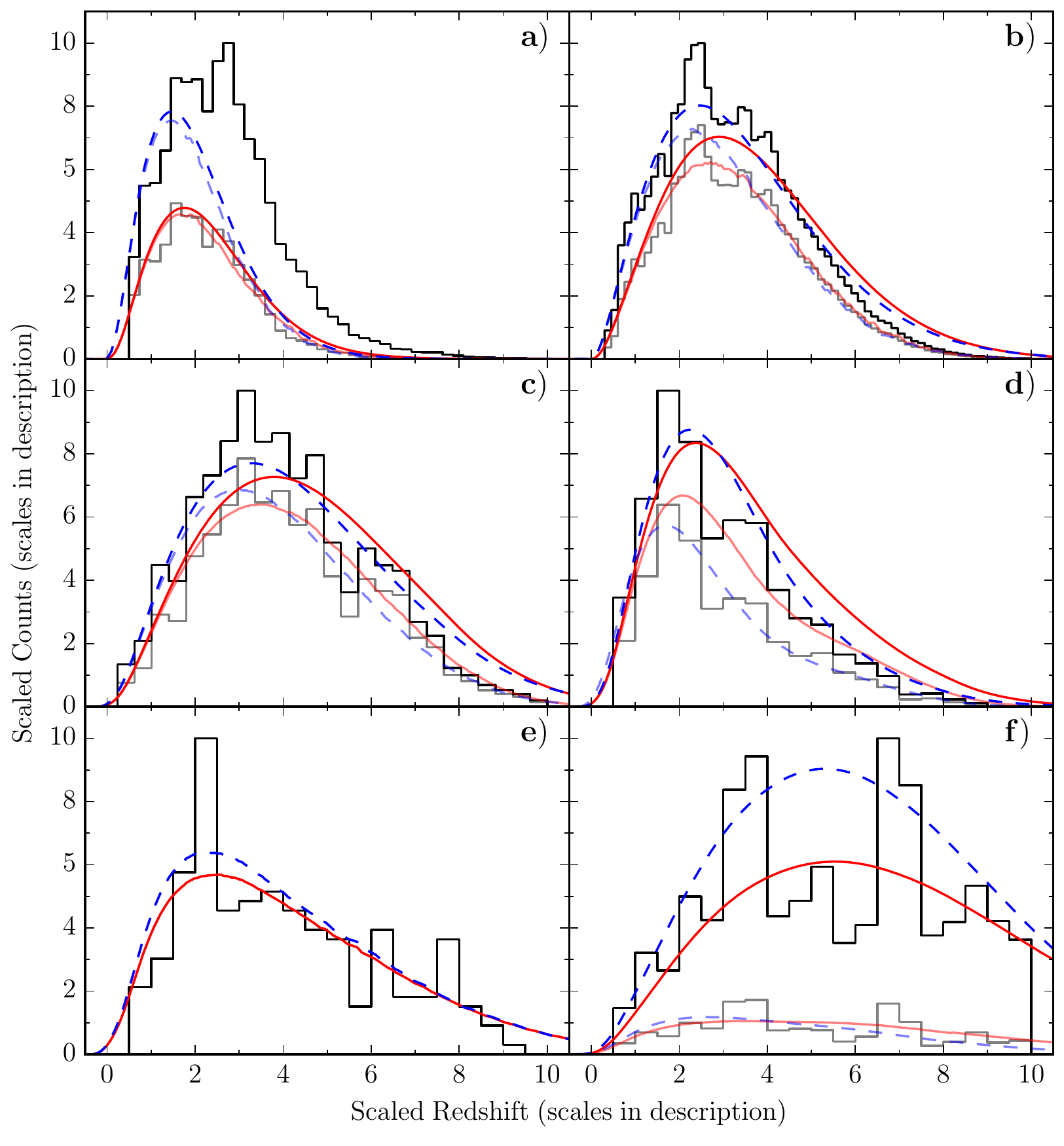}
	\end{center}
	
	\caption{Redshift Histogram Comparisons}{Above is a grid of plots comparing simple histograms of observed targets with redshift to the predicted counts from the combination of selection function and LF. 
	The grey histogram is the \WISE\ detected subset used in measuring the luminosity functions, and the black contains every high quality source in the survey. 
	The solid red lines are the predicted count density, times bin width, from the LF for the panel's survey, and the dashed blue lines are the same from the mean LF of the high-$z$ Prior Chain. 
	The light lines are models for the grey histogram, and the dark lines are for the black histogram. 
	The survey, $x$-axis scale, and $y$-axis scale for each panel, \textbf{a}--\textbf{f}, are: 6dFGS, 0.02, 432.7; SDSS, 0.033, 2922.8; GAMA, 0.043, 537.2; AGES, 0.1, 158.8; \WD, 0.1, 3.3; and zCOSMOS, 0.1, 82.1, respectively. }
	\label{fig:zhist}
\end{figure*}

It is also profitable to compare the observed redshift histograms against the predictions based on the luminosity function and selection function; plots containing such comparisons can be found in Figure~\ref{fig:zhist}. 
The power of this comparison is that, unlike the binned/unbinned LF comparisons in the earlier figures, the data in this figure need not be limited to sources with measured \WISE\ fluxes. 
Thus, the comparison between the black histogram and darker lines is not entirely one of a fit with the data it was fit to, but the extrapolation of the LF and selection function with new data. 
The first noteworthy feature of the plots in Figure~\ref{fig:zhist} is that the High $z$ Prior LF (blue dashed lines) provide more accurate extrapolations, overall, than the individual survey fits (solid red lines). 

That said, there are some features of the extrapolation in Figure~\ref{fig:zhist} that need explaining. First, the model over-predicts the number of galaxies observed at low $z$ in Panel~\textbf{a}, 6dFGS.
The likely dominant culprit there is the bias against unresolved galaxies induced by the use of the 2MASS Extended Source Catalog (2MASS XSC) to produce the target list for 6dFGS, as evidenced by the dearth of sources with $\code{w1rchi2} < 3$ relative to other surveys in Figure~\ref{fig:chivz}.
Next, the high redshift tails of of Panels \textbf{a}--\textbf{c} also disagree with the model.
Confusingly, the model under-predicts the number of galaxies in the high $z$ tail of \textbf{a}, and over-predicts the tails of \textbf{b} and \textbf{c}.
Because the high-$z$ tail of the histograms is controlled primarily by the way the selection function limits sources to those with $L>L_\star$, this is the same as saying that: one, there are more $L>L_\star$ galaxies than the model predicts; and two, that those galaxies are redder than average in the optical for the $r$ selection of SDSS and GAMA to remove them.
This appears consistent with the textbook level knowledge that the galaxies found in clusters are redder and larger than the smaller and bluer field galaxies.
Put simply, this is probably a limitation of the single SED single LF model used in this paper.
We have not performed any quantitative investigations into whether this explanation is sufficient because to do so would be to use multiple luminosity functions with their own mean SEDs to build a more accurate spectro-luminosity functional and then fit that to the data, which is beyond the scope of this work.

We have, however, investigated a number of factors that should contribute to the disagreement, but all of them are either the wrong magnitude or incapable of explaining both of the high $z$ tails: 
\begin{itemize}
\item any \WISE\ related selection effects (the extrapolated graphs remove all \WISE-based selection criteria),
\item contamination of the aperture photometry in the target selection catalogs causing sources that are too faint to be included (number of sources is too small, and makes the under-predictions worse), and
\item the spread of predictions consistent with the uncertainty in the model parameters (the $1$-$\sigma$ band is thinner than the lines in the high $z$ regions). 
\end{itemize} 
 
%


\subsection{External Comparisons}\label{sec:externalcomp}
In order to make the Schechter parameters in other papers comparable to the ones measured in this one it was necessary to use the mean SED from this work to color correct their values of $L_\star$ and, wherever possible, use the evolution measured in the other papers to bring them all to redshifts $0$, $0.38$, and $1.5$. 
The literature parameters can be found in Table~\ref{tbl:res:mainlit} alongside the mean High $z$ Prior parameters from this work. 
Further, the external works often have measurements of the LF in multiple filters; when that is the case, the observation filter with wavelength closest to the W1's $3.4\micron$ was used for the primary parameter comparisons. 
On the whole, this work's estimate of $L_\star$ is lower than the literature and its estimate for $\phi_\star$ is higher, but not radically so, especially compared to the spread among the literature values. 
The values for $L_\star$ at high redshift ($z=1.5$) have a much larger spread, making this epoch ripe for studies based on deeper imaging surveys.

The spread in measured specific evolution rates, shown in Table~\ref{tbl:res:evlit}, is considerably larger than the primary parameters. 
The uncertainties are not included in the table, but they're generally more than $0.1\,t_H^{-1}$ and less than $1\,t_H^{-1}$. 
The most directly comparable values, the ones at $^{0.1}z$, $[3.6]$, and $2.4\micron$, are all largely consistent. 
Most importantly, the specific rate of change in the density of galaxies, $R_n$, should be the same for all the different surveys in all bandpasses. 
That the spread is so large is likely attributable to a combination of selection effects, the inadequacy of a single Schechter function to describe all types of galaxies in all bandpasses, and the fact that $R_n$ is one of the least well constrained parameters by the data.

\begin{figure}[htb]
	\begin{center}
	\includegraphics[width=0.3\textwidth]{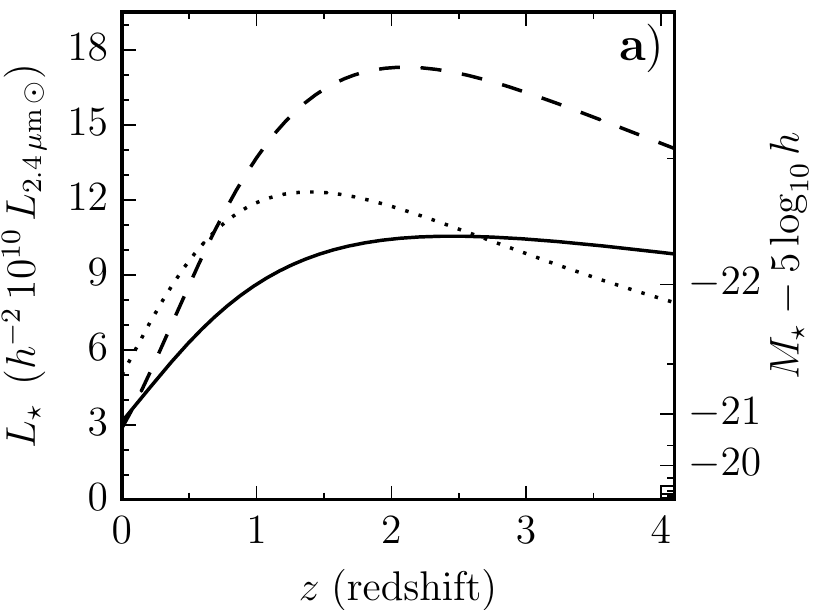}
	\includegraphics[width=0.3\textwidth]{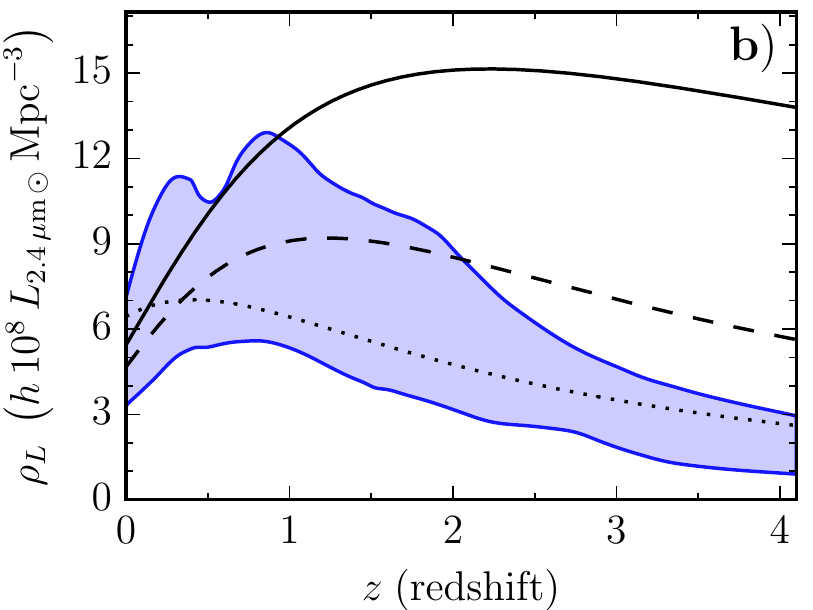}
	\includegraphics[width=0.3\textwidth]{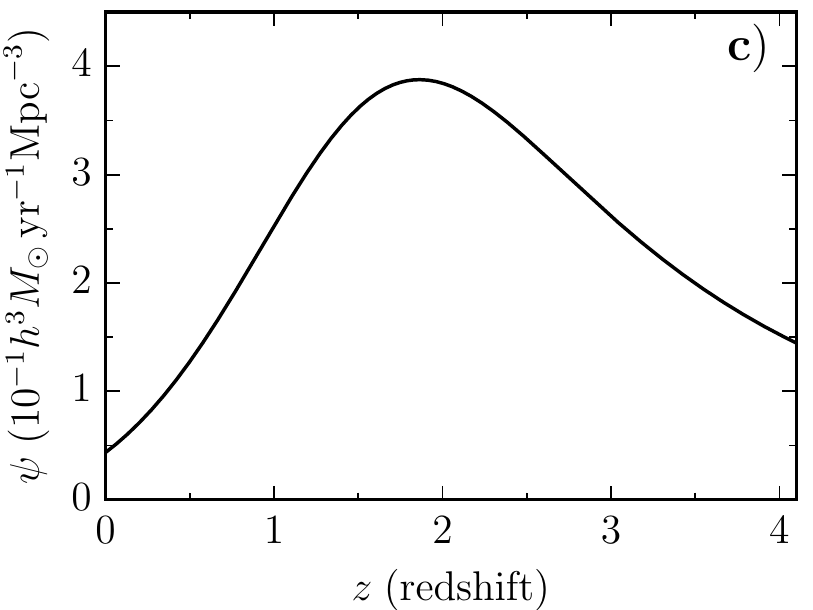}
	\end{center}
	
	\caption{Luminosity Evolution}{Graphs of empirical models for the evolution of the luminosity of galaxies. 
	Black lines in panels \textbf{a} and \textbf{b} are a plot of models from this work, the light blue region in panel \textbf{b} is a $1$-$\sigma$ confidence band from a meta-analysis in \cite{Scully:2014} \citep[$K$-band scaled with the mean SED, data primarily from][]{Arnouts:2007}, and panel \textbf{c} is the empirical model for the star formation rate density reviewed in \cite{Madau:2014}. 
	In panels \textbf{a} and \textbf{b} the identities of the lines are: the dotted line is the fit to the \WD\ sample, the dashed line is to the High $z$ Trim Prior sample, and the solid line is to the High $z$ Prior sample.}
	\label{res:fig:EVplots}
\end{figure}

The final comparison is a graphical one of the models for the evolution of $L_\star$ and $\rho_{L2.4\micron}$ from this work to an empirical models from \cite{Madau:2014} and \cite{Scully:2014} in this work's Figure~\ref{res:fig:EVplots}. 
The model from the review in \cite{Madau:2014} is for the evolution of the cosmic star formation rate density, $\psi$ (Equation~15 there), and the the model from \cite{Scully:2014} is a $1$-$\sigma$ variability band in luminosity density evolution for the  $K$ filter ($\lambda \approx 2.2\micron$) scaled using the mean SED ($L_K / L_\nu(2.4\micron) = 1.199$). 
The data used to produce the empirical model from \cite{Madau:2014} was, essentially, scaled luminosity densities measured in the far ultraviolet and far infrared. 
Most importantly, the measurements used to produce the empirical model for $\psi$ spans the range of redshifts from $0$ to $8$, so their estimate would place the peak of the luminosity density somewhere between redshifts $1.3$ and $2.5$ based on data. 
The data on $\rho_K$ evolution from \cite{Scully:2014} is primarily below redshift $2$, and based on statistically interpolating measured values from the literature, and constructing the band from an ensemble of such interpolations. 
Considering that all of the data used in this work has a redshift below $1$, and most of that less than $0.5$, the crude evolution model used here does surprisingly well at locating the epoch where $L_\star$ and $\rho_{L2.4\micron}$ peak. 
Figure~\ref{res:fig:EVplots} also shows that the peaks in $L_\star$ and $\rho_L$ are, possibly, too broad, but this is not surprising given the simplicity of the model and limits of the data used.

\floattable
\begin{deluxetable}{lccc|ccc|ccc}
	\tabletypesize{\scriptsize}
	\rotate
	\tablewidth{\textheight}
	\setlength{\tabcolsep}{4pt}
	\tablecaption{Comparison with Other IR Schechter LF Measurements}
	\tablehead{\colhead{Paper} & \colhead{Obs Band} & \colhead{$\alpha$} & \colhead{$10^{-8}\,\rho_{L2.4\micron}(0)$} & \colhead{$\phi_\star(0)$} & \colhead{$\phi_\star(0.38)$} & \colhead{$\phi_\star(1.5)$} & \colhead{$L_\star(0)$} & \colhead{$L_\star(0.38)$} & \colhead{$L_\star(1.5)$} \\
	\colhead{---} & \colhead{---} & \colhead{---} & \colhead{$h L_{2.4\micron\,\odot} \Mpc^{-3}$} &  & \colhead{$10^{-2}h^3 \Mpc^{-3}$} & & &  \colhead{$10^{10}\,L_{2.4\micron \, \odot} h^{-2}$} & }
	\startdata
		\cite{Loveday:2000} & $K$ & $-1.2\pm0.2$ & $7\pm6$ & $1.2\pm0.8$ & --- & --- & $5\pm2$ & --- & --- \\
		\cite{Kochanek:2001} & $K_s$ & $-1.09\pm0.06$ & $7.5\pm0.9$ & $1.2\pm0.1$ & --- & --- & $4.5\pm0.2$ & --- & --- \\
		\cite{Pozzetti:2003}\tablenotemark{b} & $K_s$ & $-1.2 \pm 0.1$ & $2\pm3$ & $0.1\pm0.1$ & $0.2$ & $0.4$ & $19\pm9$ & $17$ & $14$ \\
		\cite{Bell:2003} & $K_s$ & $-0.86 \pm 0.04$ & $16\pm1$ & $4.2\pm0.2$ & --- & --- & $4.1\pm0.2$ & --- & --- \\
		\cite{Babbedge:2006} & c1\tablenotemark{a} & $-0.92 \pm 0.04$ & $5\pm 5$ & $1\pm1$ & $2$ & $2$ & $4.4\pm0.4$ & $0.28$ & $0.25$ \\
		\cite{Cirasuolo:2007}\tablenotemark{b} & $K$ & $-0.99\pm0.04$ & $5.0\pm0.8$ & $0.65\pm0.09$ & $0.80$ & $0.63$ & $7.8\pm0.7$ & $7.8$ & $10.0$ \\
		\cite{Arnouts:2007}\tablenotemark{b} & $K$ & $-1.1\pm0.2$ & $12\pm3$ & $1.5\pm0.6$ & $1.1$ & $0.64$ & $8\pm3$ & $8$ & $11$ \\
		\cite{Dai:2009} & c1\tablenotemark{a} & $-1.12 \pm 0.16$ & $3.9\pm0.9$ & $1.08\pm0.03$ & --- & --- & $3.3\pm0.6$ & $5.1$ & $18$ \\
		\cite{Smith:2009} & $K$ & $-0.81\pm0.04$ & $5.7\pm0.3$ & $1.66\pm0.08$ & --- & --- & $3.7\pm0.1$ & $5.3$ & $14.9$ \\ 
		\cite{Loveday:2012} & $z$ & $-1.07\pm0.02$ & $5.1\pm0.8$ & $1.3\pm0.2$ & $1.1$ & $0.63$ & $3.9\pm0.2$ & $7.0$ & $41$ \\
		\cite{Kelvin:2014} & $K$ & $-1.16\pm0.04$ & $7\pm2$ & $0.7\pm0.1$  & --- & --- & $9\pm2$ & --- & ---   \\
		This work & W1 & $-1.050\pm 0.004$ & $5.4 \pm 0.2$ & $1.69 \pm 0.08$ & $1.6$ & $1.5$ & $3.12 \pm 0.05$ & $5.6$ & $9.9$ \\
	\enddata
	\tablecomments{Paper is the work from which the measurements came, in chronological order. 
	`Obs Band' is the observation band used to calculate the luminosity function. 
	$\rho_{L2.4\micron}(0)$ is the $2.4\micron$ luminosity density at $z=0$, $\phi_\star(z)$ is the LF normalization at redshift $z$, and $L_\star(z)$ is the value of $L_\star$ at redshift $z$. 
	Uncertainties from this work do not include systematic uncertainties, and so are underestimated. 
	The conversion factors used are: \mbox{$L_K / L_\nu(2.4\micron) = 1.199$,} \mbox{$L_{K_s} / L_\nu(2.4\micron) = 1.234$,} \mbox{$L_{\mathrm{c1}} / L_\nu(2.4\micron) = 0.842$,} and \mbox{$L_z / L_\nu(2.4\micron) = 0.924$.}}
	\tablenotetext{a}{\textit{Spitzer}/IRAC channel 1, $\lambda=3.6\micron$.}
	\tablenotetext{b}{Parameters here are based on linear interpolation/extrapolation.}
	\label{tbl:res:mainlit}
\end{deluxetable}

\begin{deluxetable}{lcc|cc|cc}
	\tabletypesize{\scriptsize}
	\tablewidth{0.68\textwidth}
	\setlength{\tabcolsep}{4.8pt}
	\tablecaption{Evolution Rate Comparisons}
	\tablehead{\colhead{Paper} & \colhead{$z_0$}  & \colhead{Band} & \colhead{$\partial \ln L_\star / \partial t$} & \colhead{$R_\phi$} & \colhead{$R_n$} & \colhead{$R_\rho$} \\
	& & & \colhead{$t_H^{-1}$} & \colhead{$t_H^{-1}$} & \colhead{$t_H^{-1}$} & \colhead{$t_H^{-1}$} }
	\startdata
		\cite{Blanton:2003LF} & $0.1$ & $^{0.1}u$ & $-3.4$ & $-2.6$ & $-2.3$ & $-5.9$ \\
		& & $^{0.1}g$ & $-1.6$ & $-0.3$ & $-0.1$ & $-1.9$ \\
		& & $^{0.1}r$ & $-1.3$ & $-0.1$ & $-0.2$ & $-1.4$ \\
		& & $^{0.1}i$ & $-1.3$ & $-0.5$ & $-0.5$ & $-1.8$ \\
		& & $^{0.1}z$ & $-0.6$ & $-1.8$ & $-1.9$ & $-2.4$ \\
		\hline
		\cite{Dai:2009} & $0.25$ & [3.6] & $-0.8$ & $0$\tablenotemark{a} & $-0.1$ & $-0.8$ \\
		& & [4.5] & $-0.7$ & $0$\tablenotemark{a} & $\hphantom{-}0.0$ & $-0.7$ \\
		\hline
		\cite{Cool:2012} & $0.3$ & $^{0.1}r$ & $-1.0$ & $\hphantom{-}0.4$ & $\hphantom{-}0.4$ & $-0.5$  \\
		\hline
		\cite{Loveday:2012} & $0.13$ & $^{0.1}u$ & $-4.8$ & $\hphantom{-}6.5$ & $\hphantom{-}6.1$ & $\hphantom{-}1.8$ \\
		& & $^{0.1}g$ & $-2.2$ & $\hphantom{-}1.2$ & $\hphantom{-}0.9$ & $-1.1$ \\
		& & $^{0.1}r$ & $-0.5$ & $-1.4$ & $-1.5$ & $-1.9$ \\
		& & $^{0.1}i$ & $-1.1$ & $\hphantom{-}0.0$ & $-0.1$ & $-1.2$ \\
		& & $^{0.1}z$ & $-1.3$ & $\hphantom{-}0.4$ & $\hphantom{-}0.3$ & $-0.9$ \\
		\hline
		\cite{Loveday:2015} & $0.2$ & $^{0.1}r$ & $-0.7$ & $-0.7$ & $-0.9$ & $-1.4$ \\
		\hline
		This work & $0.38$ & $2.4\micron$ & $-1.8$ & $\hphantom{-}0.2$ & $\hphantom{-}0.1$ & $-1.7$ \\
	\enddata
	\tablecomments{Paper is the work from which the measurements came, in chronological order. 
	$z_0$ is the mean or median redshift of the data in the work, and is the redshift at which the parameters were evaluated to make this table. 
	Band is the passband in which the luminosity function was measured. 
	$\partial \ln L_\star / \partial t$ is the specific rate of change of $L_\star$, measured in units of inverse Hubble times. 
	$R_\phi$ is the specific rate of change of $\phi_\star$ (see Equations~31 of LW17I), $R_n$ is the specific rate of change in the number density of galaxies (see Equation~\ref{eqn:specnumrate}), and $R_\rho$ is the specific rate of change in the $2.4\micron$ luminosity density (see Equation~42 of LW17I) }
	\tablenotetext{a}{This parameter was set to this value, not measured.}
	\label{tbl:res:evlit}
\end{deluxetable}

\section{Discussion} \label{sec:discussion}
The consistency of the results in this work with the literature shows that, while the goal of increasing the statistical accuracy of the LF measurements has been met, that increased accuracy has not, yet, uncovered any new facets of galaxy evolution.
The internal comparisons show that there is still room for improvement in the techniques used here. 
In particular, closer attention paid to ensuring that the full flux of resolved galaxies is measured without contamination from foreground stars will permit a further large jump in sample size. 
Improvement in the performance of numerical integration of arbitrary high dimensional Gaussian functions over rectangular regions would make spectro-luminosity functional, $\Psi$, based techniques able to analyze samples constructed using the sort of complicated color selection done for DEEP2 \citep{Newman:2013}, the SDSS luminous red galaxy sample \citep{Eisenstein:2001}, and the SDSS quasar sample \citep{Richards:2002}. 

Improvements in the form of $\Psi$ that could allow it to fit the data more closely need a little more thought to the overall approach of the analysis before implementation. 
The primary approach used in the measurement of LFs is to classify galaxies into types and measure separate LFs for each type. 
It is not obvious, but the improvements suggested in LW17I, where the total $\Psi$ is written as a sum over components, falls into the classification category. 
The way that the sum over components with $\Psi$ becomes a classification scheme, effectively, is because the fluxes measured for each source will place it closest to one mean SED. 
If we define the square distance to the SED as the exponent in $\Lsed$ then the majority of sources will most strongly affect the the LF parameters that correspond to the SED to which they are closest by that distance measure -- with a few in a boundary region dividing their influence among the terms. 
Thus, especially if the mean SEDs are fixed, the model effectively classifies the sources by which term in the total spectro-luminosity functional the source has the greatest impact on. 
The more fluxes per source brought to bear in the analysis, the sharper that divide between source types is. 

The less common approach would be to, instead of classifying galaxies, analyzing their composition. 
Think of it as the difference between deciding whether a galaxy `is a' versus how much the galaxy `has a'. 
The advantage of the composition approach is that it addresses an ambiguity not dealt with directly by the model for estimating $\Psi$ developed here: how many galaxies does each source represent? 
We know from images of local galaxies that the moderate to large galaxies have smaller galaxies in their halos (for example: the Large and Small Magellanic Clouds). 
As distance increases, the light from any satellite galaxies must, inevitably, be merged into the light from their primaries. 
The hallmark of a compositional approach is that it does not just assign probabilities to a source being in different classes, it divides the source's luminosity among them. 
So, for example, it could be possible to talk about dividing the $\Psi$ into massive stars (say O, B, and A), intermediate stars (F and G), light stars (K, M, and lighter), stellar and supernova remnants, nebular emission, and AGN. 
The primary challenge would be to figure out what form the base luminosity function, $\Phi$, of each of these should take. 
While it would seem that an increased richness of photometric data would also be required in order to analyze the composition of each galaxy, it isn't necessary to get a detailed analysis of every galaxy to get an accurate picture of the average composition of galaxies that is encoded in $\Psi$.

One question that is answerable by improving the techniques developed here, and adding data for galaxies in the redshift range of $1$ to $3$ is: which peaked first, $L_\star$ or $\rho_L$? 
The reason this question is of interest is because it encodes information about the comparative rate of star formation versus galaxy mergers; the former boosts both quantities, the latter only boosts $L_\star$. 
We would, for example, expect $L_\star$ to peak first in a universe where star formation continued in small galaxies after it and collisions slowed down for large ones. 
The converse would be the case if star formation was a relatively short epoch that cut off in most galaxies and the majority of individual galaxy luminosity growth was through accretion. 
The data  and simple evolution models used here cannot answer this question, though they hint that $L_\star$ peaked first.

Another question that will require even greater care to answer is: when did (or will) the number density of galaxies peak? 
The question is equivalent to asking when $R_n=0$. 
Though that is an equation that can, in principle, be solved using the parameters in this work, the spread in values produced are so large as to make the answer meaningless. 
It would take an evolutionary model with greater physical fidelity combined with more data to provide an answer worth examining.

\section{Conclusion}\label{sec:conclusion}
The combination of the six different redshift surveys described in LW18II, made possible by the analysis techniques derived in LW17I, has produced measurements of the Schechter LF parameters that are comparable to the literature in terms of statistical precision for $\phi_\star$, and a marked improvement for $L_\star$ and $\alpha$. 
The parameters describing the evolution of $L_\star$ and $\phi_\star$, $R_L$ and $R_\phi$, are less well constrained, but still comparable to the literature. 
Improving the constraints will require refinements in the process from end to end.

The photometry of bright, resolved and marginally resolved, galaxies requires improvements that bring them in line with the quality of our photometry for point source. 
In the ideal case, a photometric survey with sufficient sensitivity to resolve the wings of the Airy profile produced by even the largest galaxies, and the software tools needed to remove foreground contamination from stars, would guarantee that close enough to all of the light in the galaxy has been directly observed to measure accurate luminosities and colors for each whole galaxy. 

The accuracy of the spectro-luminosity functional, $\Psi[L_\nu](z)$, can be improved in the straightforward way: writing it as a sum of spectro-luminosity functionals, each with its own LF and Gaussian \Lsed. 
It may also be possible to improve the performance of $\Psi$ by a rethinking how the LF is defined -- instead of classifying galaxies into mutually exclusive categories, split into constituent parts with their own luminosity. 
The upside of a such an approach is that it naturally handles cases where multiple unresolved galaxies are contained in the same object.

Perhaps more important than increasing the fidelity of $\Psi$ is adding a model of the effective radius of galaxies so that surface brightness limits on galaxy selection can be modeled. 
Likewise, re-deriving an approximation of the estimator for the likelihood of observing an entire catalog so that it includes the effect of galaxy environment. 
Adding the effect of environment, for example using the two point function ($\xi(r)$), in the likelihood of the catalog is the most natural way to introduce cosmic variance to the process.

Finally, measuring fluxes and redshifts for even fainter galaxies using instruments like the Multi-Object Spectrometer for Infra-Red Exploration (MOSFIRE) on the Keck II telescope, and the James Webb Space Telescope (JWST), will allow for explorations of the evolution of the faint end slope of galaxies, better constrain the evolution of $\phi_\star$ and $L_\star$ with more high redshift data, and even, potentially, find a downturn in the LF at faint luminosity where galaxies and star clusters overlap as gas accreting gravitationally bound systems.

\acknowledgements

We would like to thank the \textbf{\WISE} team.\\
This publication makes use of data products from the Wide-field Infrared Survey Explorer, which is a joint project of the University of California, Los Angeles, and the Jet Propulsion Laboratory/California Institute of Technology, and NEOWISE, which is a project of the Jet Propulsion Laboratory/California Institute of Technology. WISE and NEOWISE are funded by the National Aeronautics and Space Administration.

We would like to thank the \textbf{SDSS} team.\\
Funding for SDSS-III has been provided by the Alfred P. Sloan Foundation, the Participating Institutions, the National Science Foundation, and the U.S. Department of Energy Office of Science. The SDSS-III web site is http://www.sdss3.org/.

SDSS-III is managed by the Astrophysical Research Consortium for the Participating Institutions of the SDSS-III Collaboration including the University of Arizona, the Brazilian Participation Group, Brookhaven National Laboratory, Carnegie Mellon University, University of Florida, the French Participation Group, the German Participation Group, Harvard University, the Instituto de Astrofisica de Canarias, the Michigan State/Notre Dame/JINA Participation Group, Johns Hopkins University, Lawrence Berkeley National Laboratory, Max Planck Institute for Astrophysics, Max Planck Institute for Extraterrestrial Physics, New Mexico State University, New York University, Ohio State University, Pennsylvania State University, University of Portsmouth, Princeton University, the Spanish Participation Group, University of Tokyo, University of Utah, Vanderbilt University, University of Virginia, University of Washington, and Yale University.

We would like to thank the \textbf{GAMA} team.\\
GAMA is a joint European-Australasian project based around a spectroscopic campaign using the Anglo-Australian Telescope. The GAMA input catalogue is based on data taken from the Sloan Digital Sky Survey and the UKIRT Infrared Deep Sky Survey. Complementary imaging of the GAMA regions is being obtained by a number of independent survey programs including GALEX MIS, VST KiDS, VISTA VIKING, WISE, Herschel-ATLAS, GMRT and ASKAP providing UV to radio coverage. GAMA is funded by the STFC (UK), the ARC (Australia), the AAO, and the participating institutions. The GAMA website is http://www.gama-survey.org/ .

Based on observations made with ESO Telescopes at the La Silla or Paranal Observatories under programme ID 175.A-0839.

We would like to thank the \textbf{2MASS} team.\\
This publication makes use of data products from the Two Micron All Sky Survey, which is a joint project of the University of Massachusetts and the Infrared Processing and Analysis Center/California Institute of Technology, funded by the National Aeronautics and Space Administration and the National Science Foundation.

We would like to thank the \textbf{MAST} team.\\
Some/all of the data presented in this paper were obtained from the Mikulski Archive for Space Telescopes (MAST). STScI is operated by the Association of Universities for Research in Astronomy, Inc., under NASA contract NAS5-26555. Support for MAST for non-HST data is provided by the NASA Office of Space Science via grant NNX13AC07G and by other grants and contracts.

We would like to thank the \textbf{NDWFS} team.\\
This work made use of images and/or data products provided by the NOAO Deep Wide-Field Survey (Jannuzi and Dey 1999; Jannuzi et al. 2005; Dey et al. 2005), which is supported by the National Optical Astronomy Observatory (NOAO). NOAO is operated by AURA, Inc., under a cooperative agreement with the National Science Foundation.

We would like to thank the \textbf{IPAC} team.\\
This research has made use of the NASA/ IPAC Infrared Science Archive, which is operated by the Jet Propulsion Laboratory, California Institute of Technology, under contract with the National Aeronautics and Space Administration.

We would like to thank the \textbf{GALEX} team.\\
Based on observations made with the NASA Galaxy Evolution Explorer. 
GALEX is operated for NASA by the California Institute of Technology under NASA contract NAS5-98034.

We would also like to thank the teams behind \textbf{6dFGS}, \textbf{AGES}, \textbf{zCOSMOS}, \textbf{SDWFS}, and \textbf{COSMOS}.

RJA was supported by FONDECYT grant number 1151408.

%

\bibliography{ThesisBib}

\end{document}